\documentclass[apj,twocolumn]{emulateapj}
\usepackage[alpha,z,expert]{tmsfnts}
\usepackage{color}
\usepackage{natbib}
\usepackage{amssymb}

\newcommand{\chandra}{{\it Chandra}}

\newcommand{\xmm}{{\it XMM-Newton}}
\newcommand{\planck}{{\it Planck}}
\newcommand{\wmap}{{\it WMAP}}

\newcommand{\uJybm}{\mbox{$\mu$Jy beam$^{-1}$}}
\newcommand{\ruv}{\mbox{$r_{uv}$}}

\newcommand{\cone}{ACT-CL J0022--0036}
\newcommand{\ctwo}{ACT-CL J2051+0057}
\newcommand{\cthree}{ACT-CL J2337+0016}

\newcommand{\coneshort}{ACTJ0022}
\newcommand{\ctwoshort}{ACTJ2051}
\newcommand{\cthreeshort}{A2631}
\newcommand{\cthreeshortACT}{ACTJ2337}

\slugcomment{}
\shortauthors{Reese et al.}
\shorttitle{High-Resolution SZA Observations of ACT Clusters}
\begin{document}

%%%%%%%%%%%%%%%%%%%%%%%%%%%%%%%%%%%%%%%%%%%%%%%%%%%%%%%%%%%%%%%%%%%%%%
%
%%%%%%%%%%%%%%%%%%%%%%%%%%%%%%%%%%%%%%%%%%%%%%%%%%%%%%%%%%%%%%%%%%%%%%
\title{The Atacama Cosmology Telescope: High-Resolution
  Sunyaev-Zel'dovich Array Observations of ACT SZE-selected Clusters
  from the Equatorial Strip}
%%%%%%%%%%%%%%%%%%%%%%%%%%%%%%%%%%%%%%%%%%%%%%%%%%%%%%%%%%%%%%%%%%%%%%
%%%%%%%%%%%%%%%%%%%%%%%%%%%%%%%%%%%%%%%%%%%%%%%%%%%%%%%%%%%%%%%%%%%%%%
\author{
  Erik~D.~Reese\altaffilmark{1},
  Tony~Mroczkowski\altaffilmark{1,2},
  Felipe~Menanteau\altaffilmark{3},
  Matt~Hilton\altaffilmark{4},
  Jonathan~Sievers\altaffilmark{5},
  Paula~Aguirre\altaffilmark{6},
  John~William~Appel\altaffilmark{7},
  Andrew~J.~Baker\altaffilmark{3},
  J.~Richard~Bond\altaffilmark{5},
  Sudeep~Das\altaffilmark{8,10,7}
  Mark~J.~Devlin\altaffilmark{1},
  Simon~R.~Dicker\altaffilmark{1},
  Rolando~D\"{u}nner\altaffilmark{6},
  Thomas~Essinger-Hileman\altaffilmark{7},
  Joseph~W.~Fowler\altaffilmark{9,10},
  Amir~Hajian\altaffilmark{5,10,7},
  Mark~Halpern\altaffilmark{11},
  Matthew~Hasselfield\altaffilmark{11},
  J.~Colin~Hill\altaffilmark{10},
  Adam~D.~Hincks\altaffilmark{7},
  Kevin~M.~Huffenberger\altaffilmark{12},
  John~P.~Hughes\altaffilmark{3},
  Kent~D.~Irwin\altaffilmark{9},
  Jeff~Klein\altaffilmark{1},
  Arthur~Kosowsky\altaffilmark{13},
  Yen-Ting~Lin\altaffilmark{14,6,10},
  Tobias~A.~Marriage\altaffilmark{15,10},
  Danica~Marsden\altaffilmark{1},
  Kavilan~Moodley\altaffilmark{16},
  Michael~D.~Niemack\altaffilmark{9,7},
  Michael~R.~Nolta\altaffilmark{5},
  Lyman~A.~Page\altaffilmark{7},
  Lucas~Parker\altaffilmark{7},
  Bruce~Partridge\altaffilmark{17},
  Felipe~Rojas\altaffilmark{6},
  Neelima~Sehgal\altaffilmark{18},
  Crist\'obal~Sif\'on\altaffilmark{6},
  David~N.~Spergel\altaffilmark{10},
  Suzanne~T.~Staggs\altaffilmark{7},
  Daniel~S.~Swetz\altaffilmark{1,9},
  Eric~R.~Switzer\altaffilmark{19,1},
  Robert~Thornton\altaffilmark{1,20},
  Hy~Trac\altaffilmark{21},
  Edward~J.~Wollack\altaffilmark{22}
}
%%%%%%%%%%%%%%%%%%%%%%%%%%%%%%%%%%%%%%%%%%%%%%%%%%%%%%%%%%%%%%%%%%%%%%
\altaffiltext{1}{Department of Physics and Astronomy, University of
  Pennsylvania, 209 South 33rd Street, Philadelphia, PA, 19104, USA}
\altaffiltext{2}{Einstein Postdoctoral Fellow}
\altaffiltext{3}{Department of Physics and Astronomy, Rutgers, 
  The State University of New Jersey, Piscataway, NJ 08854-8019, USA}
\altaffiltext{4}{School of Physics and Astronomy, University of
  Nottingham, University Park, Nottingham, NG7 2RD, UK} 
\altaffiltext{5}{Canadian Institute for Theoretical Astrophysics, University of
  Toronto, Toronto, ON, M5S 3H8, Canada}
\altaffiltext{6}{Departamento de Astronom{\'{i}}a y Astrof{\'{i}}sica, 
  Facultad de F{\'{i}}sica, Pontificia Universidad Cat\'{o}lica de Chile,
  Casilla 306, Santiago 22, Chile}
\altaffiltext{7}{Joseph Henry Laboratories of Physics, Jadwin Hall,
  Princeton University, Princeton, NJ, 08544, USA}
\altaffiltext{8}{Berkeley Center for Cosmological Physics, LBL and
  Department of Physics, University of California, Berkeley, CA 94720,
  USA}
\altaffiltext{9}{NIST Quantum Devices Group, 325
  Broadway Mailcode 817.03, Boulder, CO, 80305, USA}
\altaffiltext{10}{Department of Astrophysical Sciences, Peyton Hall, 
  Princeton University, Princeton, NJ 08544, USA}
\altaffiltext{11}{Department of Physics and Astronomy, University of
  British Columbia, Vancouver, BC, V6T 1Z4, Canada}
\altaffiltext{12}{Department of Physics, University of Miami, Coral Gables, 
  FL, 33124, USA}
\altaffiltext{13}{Department of Physics and Astronomy, University of
  Pittsburgh, Pittsburgh, PA, 15260, USA}
\altaffiltext{14}{Institute for the Physics and Mathematics of the Universe, 
  The University of Tokyo, Kashiwa, Chiba 277-8568, Japan} 
\altaffiltext{15}{Dept. of Physics and Astronomy, The Johns Hopkins
  University, 3400 N. Charles St., Baltimore, MD 21218-2686, USA} 
\altaffiltext{16}{Astrophysics and Cosmology Research Unit, School of
  Mathematical Sciences, University of KwaZulu-Natal, Durban, 4041,
  South Africa}
\altaffiltext{17}{Department of Physics and Astronomy, Haverford
  College, Haverford, PA, 19041, USA} 
\altaffiltext{18}{Kavli Institute for Particle Astrophysics and
  Cosmology, Stanford University, Stanford, CA, 94305-4085, USA}
\altaffiltext{19}{Kavli Institute for Cosmological Physics, 
  Laboratory for Astrophysics and Space Research, 5620 South Ellis Ave.,
  Chicago, IL, 60637, USA}
\altaffiltext{20}{Department of Physics , West Chester University 
  of Pennsylvania, West Chester, PA, 19383, USA}
\altaffiltext{21}{Department of Physics, Carnegie Mellon University,
  Pittsburgh, PA 15213, USA} 
\altaffiltext{22}{Code 553/665, NASA/Goddard Space Flight Center,
  Greenbelt, MD, 20771, USA}
%\email{erreese@physics.upenn.edu}
%%%%%%%%%%%%%%%%%%%%%%%%%%%%%%%%%%%%%%%%%%%%%%%%%%%%%%%%%%%%%%%%%%%%%%
\begin{abstract}
We present follow-up observations with the Sunyaev-Zel'dovich Array
(SZA) of optically-confirmed galaxy clusters found in the equatorial
survey region of the Atacama Cosmology Telescope (ACT): \cone, \ctwo,
and \cthree.  \cone\ is a newly-discovered, massive ($\simeq
10^{15}$~$M_\odot$), high-redshift ($z=0.81$) cluster revealed by ACT
through the Sunyaev-Zel'dovich effect (SZE).  Deep, targeted
observations with the SZA allow us to probe a broader range of cluster
spatial scales, better disentangle cluster decrements from radio point
source emission, and derive more robust integrated SZE flux and mass
estimates than we can with ACT data alone.  For the two clusters we
detect with the SZA we compute integrated SZE signal and derive masses
from the SZA data only.  \cthree, also known as Abell 2631, has
archival \chandra\ data that allow an additional X-ray-based mass
estimate.  Optical richness is also used to estimate cluster masses
and shows good agreement with the SZE and X-ray-based estimates.
Based on the point sources detected by the SZA in these three cluster
fields and an extrapolation to ACT's frequency, we estimate that point
sources could be contaminating the SZE decrement at the $\lesssim
20$\% level for some fraction of clusters.
\end{abstract}
%%%%%%%%%%%%%%%%%%%%%%%%%%%%%%%%%%%%%%%%%%%%%%%%%%%%%%%%%%%%%%%%%%%%%%

%%%%%%%%%%%%%%%%%%%%%%%%%%%%%%%%%%%%%%%%%%%%%%%%%%%%%%%%%%%%%%%%%%%%%%
\section{Introduction}
\label{sec:intro}
%%%%%%%%%%%%%%%%%%%%%%%%%%%%%%%%%%%%%%%%%%%%%%%%%%%%%%%%%%%%%%%%%%%%%%

The Sunyaev-Zel'dovich effect (SZE) is a small (typically $\lesssim
1$~mK) distortion of the cosmic microwave background (CMB) spectrum
caused by the inverse Compton scattering of CMB photons off energetic
electrons of the hot intracluster medium of galaxy clusters
(\citealt{zeldovich1969, sunyaev1970, sunyaev1972}; for reviews see,
\citealt{sunyaev1980, rephaeli1995, birkinshaw1999, carlstrom2002}).
The redshift independence of the SZE makes it a potentially powerful
tool with which to search for galaxy clusters, especially in the
distant universe.  Abundances of clusters probe the growth of
structure and have placed useful constraints on the fluctuation
amplitude, $\sigma_8$, and the matter density, $\Omega_M$
\citep[e.g.,][]{henry1991, viana1996, bahcall1997, eke1998,
  borgani2001, reiprich2002, schuecker2003, henry2004, mantz2008,
  vikhlinin2009b, mantz2010, rozo2010, vanderlinde2010, sehgal2011}.
The evolution of cluster abundance with redshift is one of the few
probes of the growth of structure and has the potential to tightly
constrain cosmological parameters and provide insight into the
equation of state of the dark energy \citep[e.g.,][]{bartlett1994,
  holder2000, haiman2001, majumdar2004}.  The biggest challenge to
realizing the cosmological potential of cluster surveys is relating
cluster mass to an observable such as integrated SZE signal.
Well-determined masses for even a subsample of clusters can
significantly improve parameter constraints
\cite[e.g.,][]{majumdar2004}.

The last few years have seen significant advances in surveys using the
SZE.  The Atacama Cosmology Telescope \citep[ACT;][]{fowler2007}, the
South Pole Telescope \citep[SPT;][]{carlstrom2011}, and the
\planck\ satellite \citep{planck2011a} are, for the first time,
producing catalogs of galaxy clusters discovered through the SZE
\citep{vanderlinde2010, marriage2011b, planck2011d, williamson2011}.
The fast mapping speeds and improved sensitivities of these
instruments have enabled them to survey large regions of the sky with
sufficient depth to detect massive clusters, but their limited angular
resolutions ($\gtrsim 1\arcmin$) do not allow detailed studies of
cluster astrophysics with their data alone.

We have performed initial follow-up observations with the
Sunyaev-Zel'dovich Array (SZA) of three optically-confirmed ACT
clusters in ACT's equatorial strip.  Deep, targeted SZA observations
of ACT clusters provide higher sensitivity over a broader range of
cluster spatial scales than that of ACT.  In addition, the spatial
filtering of the interferometer provides a method of cleanly
disentangling radio point source emission from the cluster SZE signal.
ACT observational details and target selection are presented in
Section~\ref{sec:target_select}. SZA and \chandra\ observations and
data reduction are described in Sections~\ref{subsec:sza_obs} and
\ref{subsec:chandra_obs}, respectively.  The analysis method and
results are reported in Section~\ref{sec:analysis}, the implications
in Section~\ref{sec:discussion}, and conclusions in
Section~\ref{sec:conclusion}.

Throughout this paper, all uncertainties are reported at 68\%
confidence and we adopt a flat, $\Lambda$-dominated cosmology with
$\Omega_M = 0.3$, $\Omega_\Lambda = 0.7$, and $H_0 = 70$ km s$^{-1}$
Mpc$^{-1}$ consistent with recent {\it Wilkinson Microwave Anisotropy
  Probe} (\wmap) results \citep{komatsu2011, komatsu2009}.

%%%%%%%%%%%%%%%%%%%%%%%%%%%%%%%%%%%%%%%%%%%%%%%%%%%%%%%%%%%%%%%%%%%%%%%
\section{ACT Target Selection}
\label{sec:target_select}
%%%%%%%%%%%%%%%%%%%%%%%%%%%%%%%%%%%%%%%%%%%%%%%%%%%%%%%%%%%%%%%%%%%%%%%

The Atacama Cosmology Telescope (ACT) is a 6-m diameter telescope
located at an elevation of 5200~m in the Atacama desert of Chile.
Three 1024-element transition-edge-sensing bolometer arrays operate at
148, 218, and 277~GHz and survey large regions of the sky mainly in
two regions, a southern strip centered at declination $-52.5^\circ$
and in an equatorial strip that encompasses the Sloan Digital Sky
Survey (SDSS) Stripe 82 \citep[hereafter S82,][]{abazajian2009}. For
the ACT instrument, observation, reduction, and calibration details
see \citet{fowler2007}, \citet{swetz2011}, \citet{das2011}, and
\citet{hajian2011}.  Initial results from the ACT southern strip
include CMB power spectra \citep{fowler2010, das2011}, compact source
\citep{marriage2011} and cluster catalogs \citep{marriage2011b},
cluster follow-up \citep{menanteau2010}, and analysis of the
cosmological implications of CMB power spectra \citep{dunkley2011} and
cluster yields \citep{sehgal2011}.

A matched filter method \citep{haehnelt1996, herranz2002, melin2006}
is used for cluster detection to produce a cluster catalog \citep[for
  ACT-specific details, see][]{marriage2011b}.  We focus on the S82
region as optical data are available from which we measure cluster
redshifts and estimate cluster masses via richness and luminosity.  We
chose three preliminary high signal-to-noise (S/N$\approx 5$) cluster
detections from the S82 region for follow-up as a pilot study for SZA
observations; these were \cone, \ctwo, and \cthree.  We will refer to
these clusters as \coneshort, \ctwoshort, and \cthreeshort, where
\cthree\ is a previously known Abell cluster.  The upper row of panels
of Figure~\ref{fig:low_res} shows the SZE images of these three fields
expressed as Compton-$y$ (see Eq.~\ref{eq:sze_defined}) made from
recent ACT maps.  \coneshort\ is a new high-redshift cluster found by
ACT through the SZE.  \coneshort\ was spectroscopically confirmed at
$z=0.81$ from long slit observations of the brightest cluster galaxy
(BCG) conducted at the Apache Point and Gemini South Observatories
(more details on these observations will be described in Menanteau
et~al. 2011, in prep., where we present the X-ray and optical
properties of the S82 ACT clusters).  The S82 data provide the
spectroscopic redshift, $0.33$, for \ctwoshort\ and \cthreeshort\ has
a spectroscopic redshift of $0.27$ \citep{struble1999}.  New maps
produced with the full ACT equatorial data set reveal \ctwoshort\ to
be lower significance (S/N $\simeq$ 4) than inferred from the
preliminary maps.  Table~\ref{tab:imstat} lists the clusters,
alternate names, and redshifts.

%%%%%
%
% Basic SZA data properties
%
%%%%%

%% rms info for OVRO/BIMA data sets
%%
%%	             low res	    high res
%%               ----------------  ---------
%% cluster,obs,  rms, beam, rms_K, rms, beam
%%
%%

\begin{deluxetable*}{lccccccccccc}

%\singlespace
%\footnotesize
%\rotate
\tabletypesize{\small}
\tablewidth{0pt}
%\tablenum{}
\tablecolumns{10}
%\tableheadfrac{}
\tablecaption{SZE Image Statistics\label{tab:imstat}}
\tablehead{
\colhead{} &
\colhead{} &
\colhead{} &
\colhead{} &
\colhead{} &
\multicolumn{3}{c}{Low Resolution ($r_{uv}<2$ k$\lambda$)} &
\colhead{} &
\multicolumn{2}{c}{High Resolution ($r_{uv}>2$ k$\lambda$)} 
\\
\cline{6-8}\cline{10-11}
\colhead{} &
\colhead{} &
\colhead{} &
\colhead{} &
\colhead{$t_{\mbox{\rm int}}$\tablenotemark{b}} &
\colhead{$\sigma$} &
\colhead{Beam} &
\colhead{$\sigma_{\mbox{\tiny CMB}}$} &
%\colhead{$\sigma_{\mbox{\tiny RJ}}$} &
\colhead{} &
\colhead{$\sigma$} &
\colhead{Beam}
\\
\colhead{Cluster} &
\colhead{Other Name} &
\colhead{$z$} &
\colhead{Timing\tablenotemark{a}} &
\colhead{(hr)} &
\colhead{(\uJybm)} &
\colhead{($^{\prime\prime} \times ^{\prime\prime}$)} &
\colhead{($\mu$K)} &
\colhead{} &
\colhead{(\uJybm)} &
\colhead{($^{\prime\prime} \times ^{\prime\prime}$)}
}
\startdata
\cone & &0.81 & pre
        & 15.7
	& 240 
	& $94 \times 107$
	& 38
	&
	& 230
	& $17 \times 20$
\\
\cone & && post
        & 21.8
	& 220 %(160)
	& $91 \times 101$
	& 38
	&
	& 230 %(170)
	& $16 \times 20$
\\
\ctwo & &0.33& pre
        & 6.7
	& 350
	& $89 \times 118$
	& 52
	&
	& 300
	& $15 \times 22$
\\
\ctwo  & && post
        & 26.9
	& 230 %(180)
	& $90 \times 105$
	& 38
	&
	& 230 %(170)
	& $16 \times 20$
\\
\cthree & Abell 2631 & 0.27& post
        & 26.6
	& 240 %(160)
	& $97 \times 106$
	& 37
	&
	& 210 %(160)
	& $17 \times 19$
%\\
\enddata
\tablenotetext{a}{Pre- and post-CARMA transition of the integration of
  SZA with CARMA.}
\tablenotetext{b}{Effective on-source integration time accounting for
  excised data.} 
\end{deluxetable*}

%%%%%%%%%%%%%%%%%%%%%%%%%%%%%%%%%%%%%%%%%%%%%%%%%%%%%%%%%%%%%%%%%%%%%%
\section{Data}
\label{sec:data}
%%%%%%%%%%%%%%%%%%%%%%%%%%%%%%%%%%%%%%%%%%%%%%%%%%%%%%%%%%%%%%%%%%%%%%

%%%%%%%%%%%%%%%%%%%%%%%%%%%%%%%%%%%%%%%%%%%%%%%%%%%%%%%%%%%%%%%%%%%%%%
\subsection{SZA Observations}
\label{subsec:sza_obs}
%%%%%%%%%%%%%%%%%%%%%%%%%%%%%%%%%%%%%%%%%%%%%%%%%%%%%%%%%%%%%%%%%%%%%%

The Sunyaev-Zel'dovich Array (SZA) is an 8-element interferometer with
3.5-m diameter dishes operating at 30 and 90~GHz.  It has an 8~GHz
bandwidth, and is located at an altitude of 2200 m in the Inyo
Mountains of California.  Our observations use the 30~GHz system,
which has a $10.\!\arcmin5$ FWHM primary beam (field of view).
Typical system temperatures are $T_{\mathrm{sys}} \simeq 40$--$50$~K,
including atmospheric contributions. Six of the antennas are placed in
a compact configuration to maximize cluster sensitivity, while two are
deployed at longer baselines to provide higher resolution data.  The
spatial filtering of the interferometer allows one to disentangle the
small-scale, positive point source emission from the large-scale,
negative SZE signal at these frequencies \citep[for an example in a
  similar context see][]{reese2002}.  The SZA has become part of the
Combined Array for Research in Millimeter-wave Astronomy (CARMA) and
our data were taken both before and after the CARMA transition.

The SZA observed the three ACT clusters between March and May and
between August and September 2010.  A bright quasar was observed every
30 minutes, for about 2 minutes integration time, to monitor the
system gain and phase during each observation.  Since the fluxes of
quasars can be variable on timescales of days or months, these
sources only serve as secondary calibrators.  We discuss absolute
calibration in Section~\ref{subsec:sza_cal}.

Pre-CARMA data are reduced using a suite of MATLAB routines developed
by the SZA collaboration \citep[e.g.,][]{muchovej2007}.  Post-CARMA
data are reduced using Miriad \citep{sault1995}.  Data are excised
when one telescope shadows another, when cluster field data are not
straddled by two phase calibrator observations, when there are
anomalous changes in instrumental response between calibrator
observations, when the system temperature changes dramatically between
integrations, or when there is spurious correlation.

%%%%%
% Low resolution maps of all three fields.
%%%%%

%%%%%%%%%%%%%%%%%%%%%%%%%%%%%%%%%%%%%%%%%%%%%%%%%%%%%%%%%%%%%%%%%%%%%%%
\begin{figure*}[!tbh]
  \centerline{
    \includegraphics[height=2.5in]{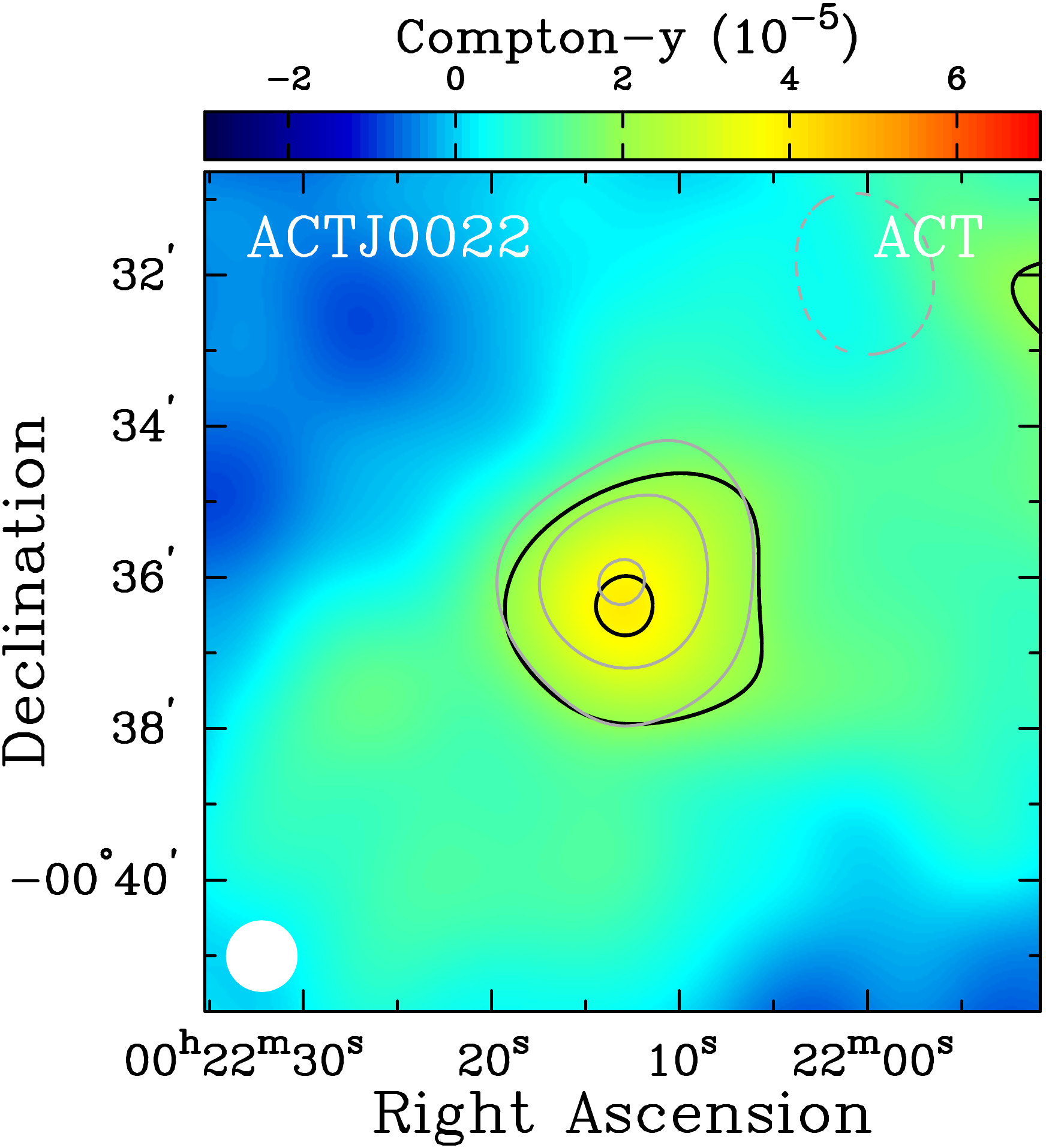}
    \includegraphics[height=2.5in]{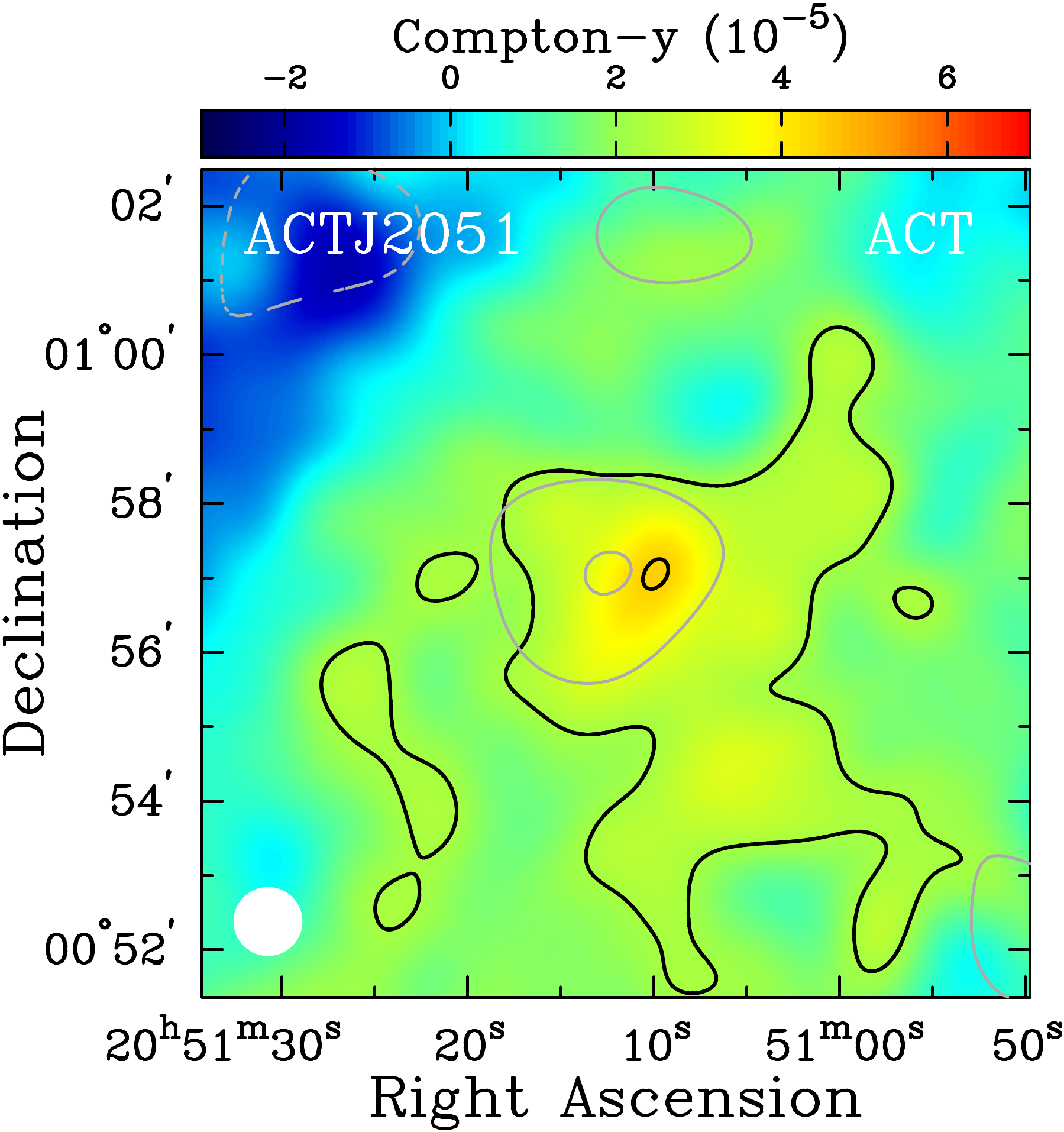}
    \includegraphics[height=2.5in]{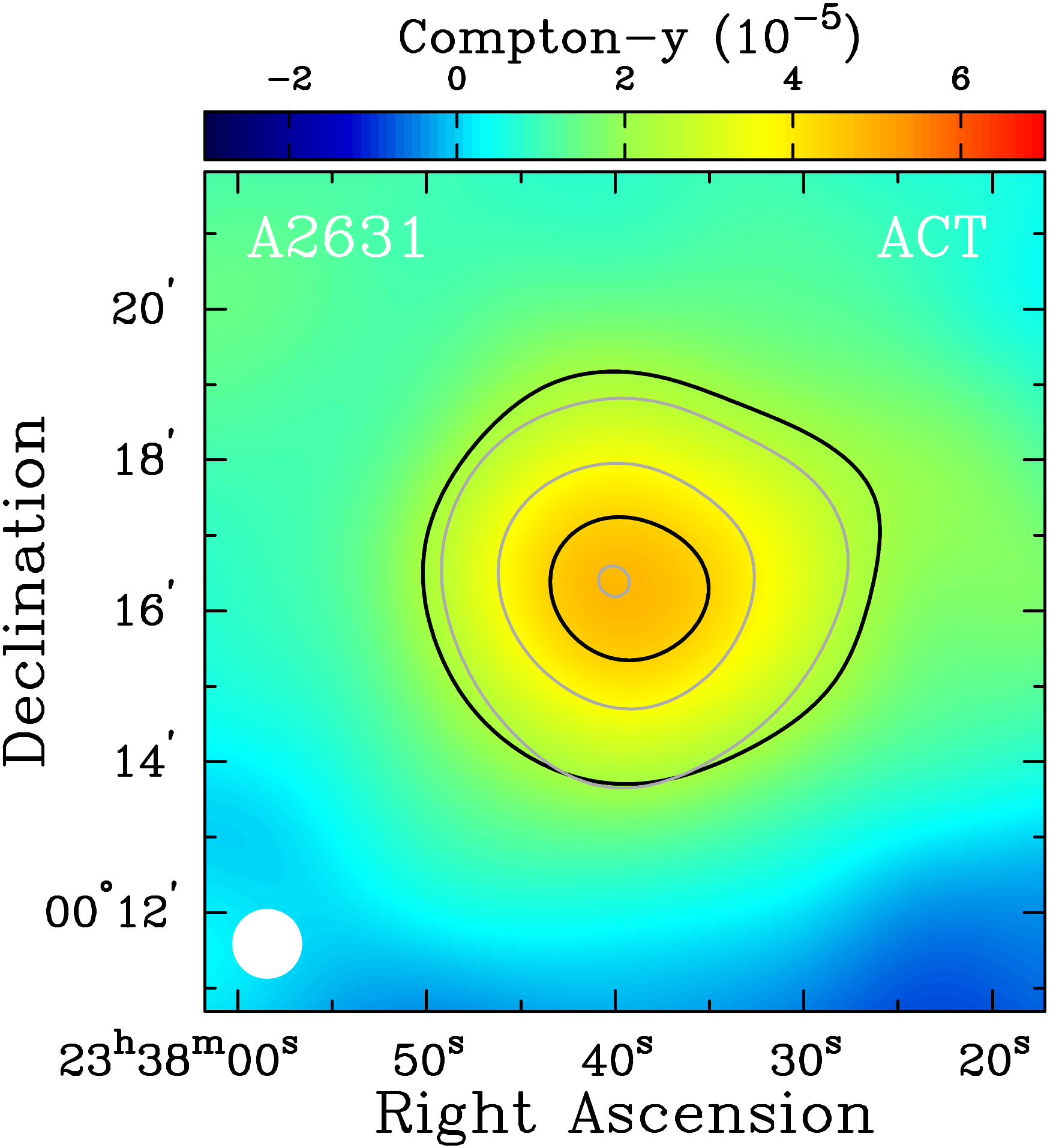}
  }
  \centerline{
    \includegraphics[height=2.5in]{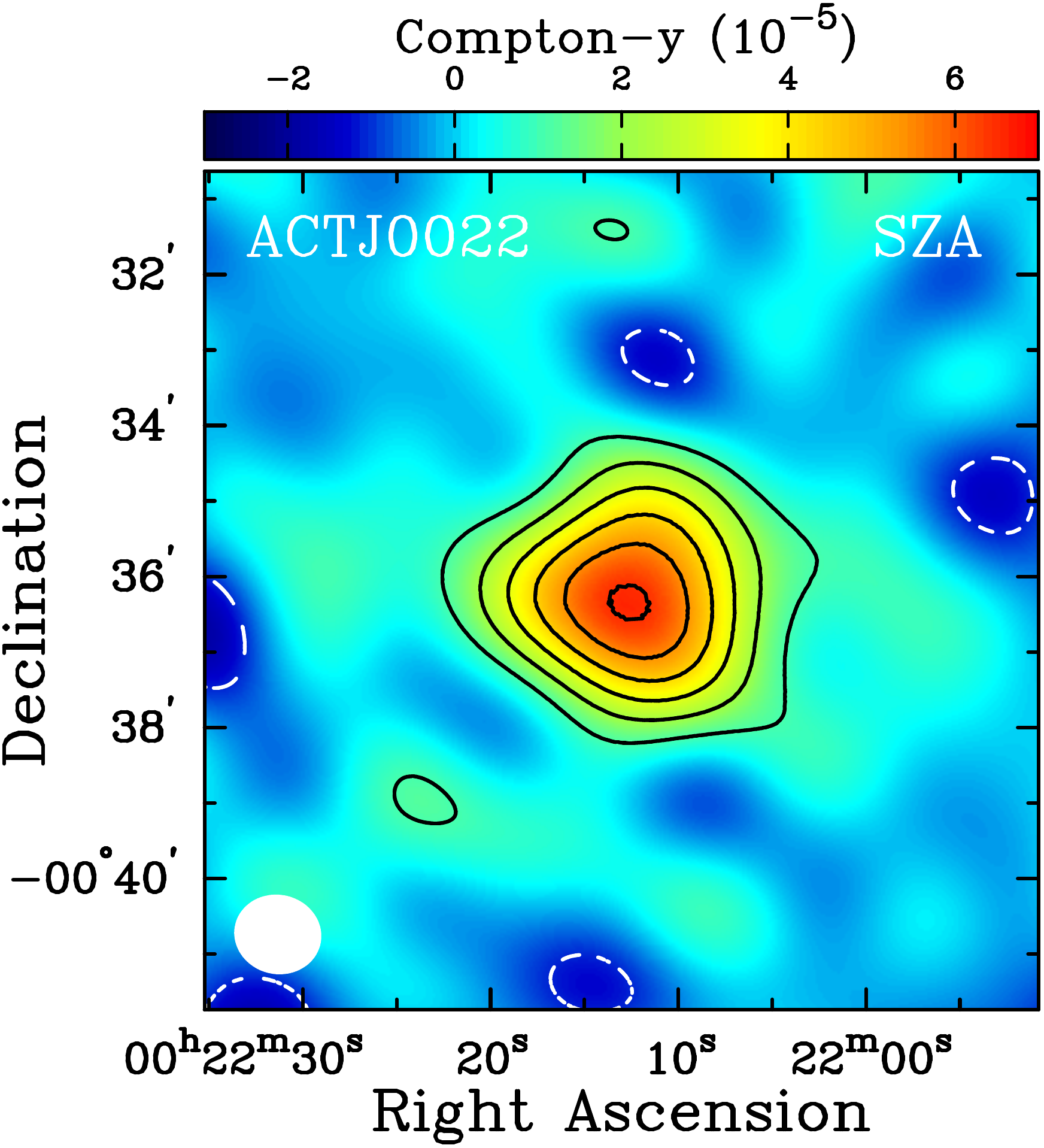}
    \includegraphics[height=2.5in]{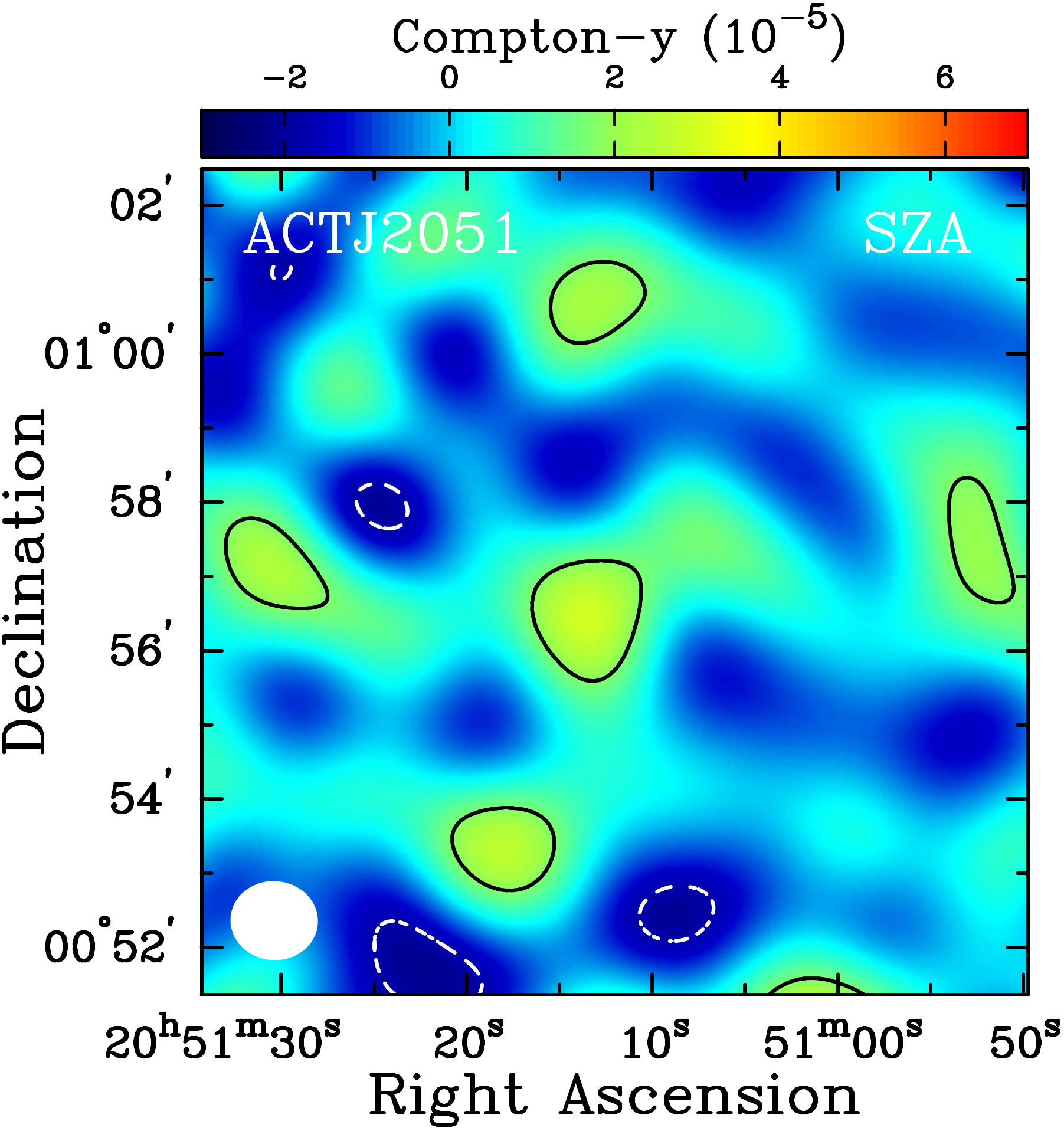}
    \includegraphics[height=2.5in]{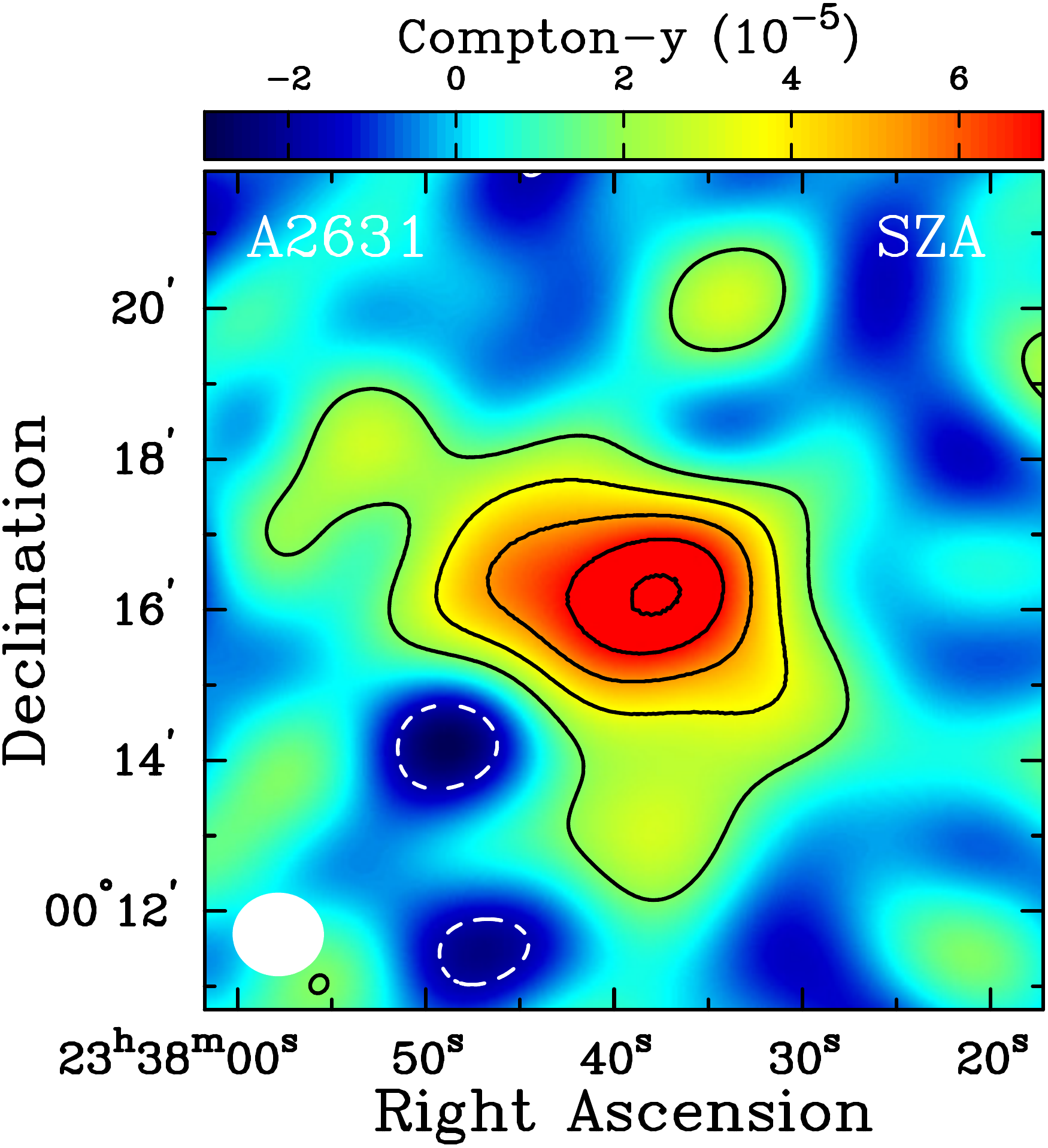}
  }
  \caption{{\it Upper}: ACT Compton-$y$ images of the three optically
    confirmed clusters made from recent ACT maps.  We fit (excluding
    the cluster region in the fit) and remove a quadratic polynomial from the
    data, smooth with a Gaussian, and resample to smaller pixels.  The
    contours are multiples of twice the rms of each map (black).  Also
    shown are contours from the match filtered map (gray) as multiples
    of twice the rms.  {\it Lower}: Deconvolved SZA images of the
    three cluster fields in units of Compton-$y$.  The SZA images are
    made with short baseline data $<2$ k$\lambda$ after removal of
    point sources.  Contours are multiples of twice the rms of each
    map (for details of the SZA observations presented here, see
    Table~\ref{tab:imstat}).  The color scales are in units of $
    10^{-5}$ Compton-$y$, and solid (dashed) lines represent positive
    (negative) signal.  All figures cover the same angular scale
    ($11\arcmin$ on a side) and are on the same color scale to
    facilitate comparison.  The FWHM of the effective resolution
    (synthesized beam for the SZA) is shown in the lower left corner
    of each panel.  For visualization purposes the SZA images
    incorporate only the low-resolution SZA data and the effective
    beams reflect this choice.  The SZA data analysis
    (Section~\ref{sec:analysis}) uses the data in its entirety.
    \coneshort\ and A2631 both yield high S/N detections.
    \ctwoshort\ is not detected by the SZA, which is consistent with
    its lower SZE signal in the current ACT maps (shown here) and with
    that expected from optical data (see
    Section~\ref{sec:discussion}).  In general, the qualitative
    agreement between the ACT and SZA data is good.
%    The SZA data provide higher resolution than ACT, and therefore
%    have higher peak Compton-$y$ values.
    \label{fig:low_res}}
\end{figure*}
%%%%%%%%%%%%%%%%%%%%%%%%%%%%%%%%%%%%%%%%%%%%%%%%%%%%%%%%%%%%%%%%%%%%%%%

The Fourier plane is also known as the $u$-$v$ plane, where $(u,v)$
are the Fourier conjugate variables to right ascension and
declination, respectively.  We define the radius in the $u$-$v$ plane
as $r_{uv}=\sqrt{u^2+v^2}$, which corresponds to the baseline length
in units of the observation wavelength, $\lambda$.  The configurations
used for these observations produce a break in $u$-$v$ coverage at 2
k$\lambda$, with similar sensitivities for the low ($r_{uv} <2$
k$\lambda$) and high ($r_{uv} >2$ k$\lambda$) resolution data (15
short and 13 long baselines).  Table~\ref{tab:imstat} summarizes the
on-source integration times, and (for the high and low-resolution data
separately) sensitivities and the FWHMs of the synthesized beams
(effective resolution).  The rms map noise, $\sigma_{\mbox{\tiny
    CMB}}$, in CMB temperature units for the given synthesized beams,
is also given for the low-resolution maps.  The total integration
times reported correspond to the equivalent time that the full
8-element, 8~GHz bandwidth SZA spent on each cluster (accounting for
excised data).

Data visualization is done with Difmap \citep{shepherd1997}.
Identification of point sources in the field is performed with the
high-resolution ($\ruv >2$~k$\lambda$) data.  Point sources are
modeled to remove them from the data, and the low-resolution data
($\ruv < 2$~k$\lambda$) are then deconvolved to produce images of the
cluster fields with sources subtracted.  We present deconvolved images
of the point source-removed SZA observations as a measure of the data
quality.  The model fitting is performed in the Fourier plane
directly.  The cluster and any point sources in the field are modeled
simultaneously using the data in its entirety.  Modeling of the
cluster and point sources is discussed in Section~\ref{subsec:method}
and results are discussed in Section~\ref{subsec:results}.

The lower panels in Figure~\ref{fig:low_res} show the deconvolved SZA
images of the three cluster fields after removal of the point sources.
Contours are multiples of twice the rms noise in each map, and the
color scale is in units of Compton-$y$.  The FWHM of the synthesized
beam (effective resolution) is shown in the lower left of each panel.
The images all cover the same angular scale ($11\arcmin$ on a side)
and have the same color scale to facilitate comparison.  The
\coneshort\ and \cthreeshort\ fields reveal high-significance SZA
detections.  The point sources in the \ctwoshort\ field are used
both to strengthen our calibration of the SZA data and to better assess
SZE decrement contamination due to such sources.

%%%%%%%%%%%%%%%%%%%%%%%%%%%%%%%%%%%%%%%%%%%%%%%%%%%%%%%%%%%%%%%%%%%%%%%
\subsection{SZA Calibration}
\label{subsec:sza_cal}
%%%%%%%%%%%%%%%%%%%%%%%%%%%%%%%%%%%%%%%%%%%%%%%%%%%%%%%%%%%%%%%%%%%%%%%

%%%%%
%
% Calibration ratios
%
%%%%%

%% Field source  Flux(pre)  Flux(post)  Ratio

\begin{deluxetable}{lccc}

%\singlespace
%\footnotesize
%\rotate
\tablewidth{0pt}
%\tablenum{}
\tablecolumns{4}
%\tableheadfrac{}
\tablecaption{Pre- and post-CARMA Point Source Flux Densities\label{tab:cal}}
\tablehead{
\colhead{} &
\colhead{$F_\nu^{\mbox{\tiny pre}}$} &
\colhead{$F_\nu^{\mbox{\tiny post}}$} &
\colhead{} 
\\
\colhead{Field} &
\colhead{(mJy)} &
\colhead{(mJy)} &
\colhead{Ratio} 
}
\startdata
\coneshort & 
$0.95 \pm 0.20$ & $0.80 \pm 0.14$ & $1.19 \pm 0.30$ 
\\                                                       
& $1.07 \pm 0.18$ & $0.65 \pm 0.13$ & $1.65 \pm 0.36$ 
\\
\ctwoshort &
$8.88 \pm 0.33$ & $5.03 \pm 0.17$ & $1.77 \pm 0.32$ 
\\
& $3.66 \pm 1.36$ & $2.49 \pm 0.61$ & $1.47 \pm 1.05$ 
\\
%\hline
%\coneshort & cluster & 
%$0.92 \pm 0.11$ & $0.61 \pm 0.08$ & $1.51 \pm 0.22$ 
%\\                                           
\hline
\\
[-0.5pc]
%Avg & & & $1.52 \pm 0.05$
%\\            
Weighted Average & & & $1.51 \pm 0.07$
%\\            
\enddata
%\tablenotetext{a}{}
%\tablecomment{}
\end{deluxetable}

Pre-CARMA SZA operations routinely monitored Mars for absolute
calibration \citep[e.g.,][]{muchovej2007}.  In the period right after
the SZA-CARMA merger, standard calibration protocols had not yet been
implemented.  Therefore, though the system gains are stable, the
absolute calibration is off by an arbitrary multiplicative factor.  We
calibrate the transitional post-CARMA data using radio sources
observed both before and after the integration to derive an average
calibration factor that is applied to the data.
This is non-ideal because the sources can vary; however, we note that
on average we do not expect the source flux densities to exhibit any
particular trend upwards or downwards over time, so the average ratio
of several sources pre- and post-CARMA may be expected to follow any
changes in calibration.  The cluster signal will remain constant, so
we can validate our calibration by fitting the same model to the pre-
and post-CARMA data and comparing the central Compton-$y$ values.

%%%%%
% Chandra-SZA overlay for A2631
%%%%%

%%%%%%%%%%%%%%%%%%%%%%%%%%%%%%%%%%%%%%%%%%%%%%%%%%%%%%%%%%%%%%%%%%%%%%%
\begin{figure}[!tbh]
  \centerline{
    \includegraphics[width=3in]{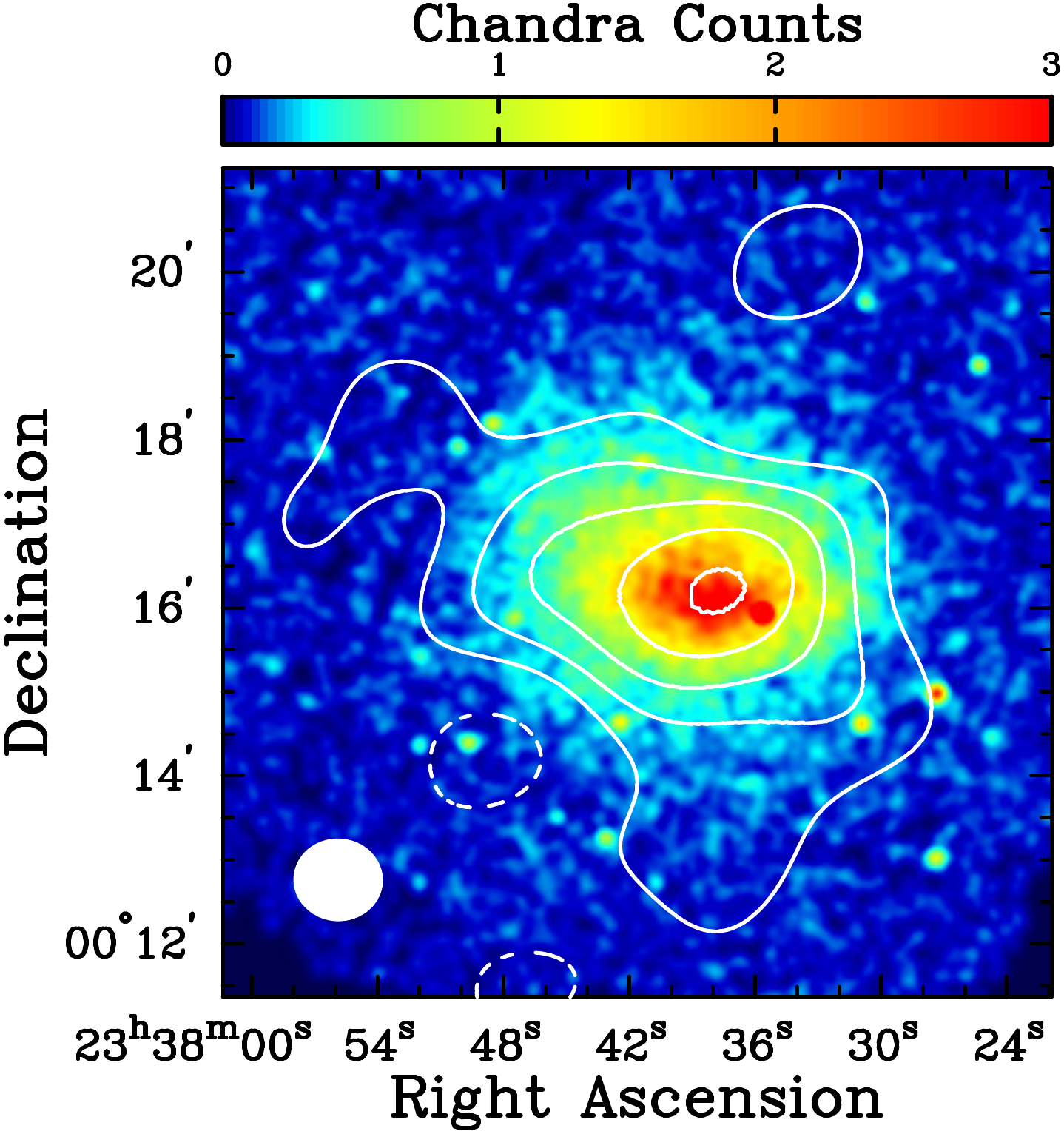}
  }
  \caption{Deconvolved SZA image of A2631 (contours) on the smoothed
    \chandra\ counts image (color).  Contours are multiples of twice
    the rms of the map (see Table~\ref{tab:imstat} for details) and
    the color scale is in units of \chandra\ counts.  The FWHM of the
    synthesized beam (effective resolution) is shown in the lower left
    corner.  The E-W elongation is seen in both the SZA and
    \chandra\ data, and the peaks in the SZE and X-ray data align
    well.
    \label{fig:a2631_x-sze}}
\end{figure}
%%%%%%%%%%%%%%%%%%%%%%%%%%%%%%%%%%%%%%%%%%%%%%%%%%%%%%%%%%%%%%%%%%%%%%%

The \coneshort\ and \ctwoshort\ fields have both pre- and post-CARMA
data.  For the calibration analysis, point source models are fit to
the data.  A cluster model is simultaneously fit to the
\coneshort\ data.  A Markov Chain Monte Carlo (MCMC) analysis is
performed to derive flux densities and uncertainties (model and fitting
details are described in Section~\ref{subsec:method}).  Ratios of pre-
to post-CARMA flux densities are computed.  The flux densities and
ratios are summarized in Table~\ref{tab:cal}.  The average ratio is
$1.5 \pm 0.1$, computed with an inverse variance weighting.  There is
a wide source-to-source variation in the ratios, which range between
1.2 and 2.0, though the uncertainties on the ratios are relatively
large and the largest discrepancy is $< 2\sigma$.  Our average ratio
is consistent with the rescaling ($\approx 1.4$) found by members of
the CARMA collaboration (T.~Plagge, private communication).

The cluster SZE signal will not change, but comparisons are
complicated by its extended structure.  We also fit the same
$\beta$-model (fixed core radius and power law index) to the pre- and
post-CARMA data for \coneshort.  The ratio of the central Compton-$y$
parameters also yields the same factor of 1.5.  That the cluster
normalizations yield the same calibration scaling as the point sources
lends some support to our simple approach.  As a result, the
post-CARMA data are simply scaled by 1.5 and used for the analysis.
Note that the rms values in Table~\ref{tab:imstat} correspond to the
scaled post-CARMA data.  The pre-CARMA data absolute flux calibration
is known to better than 10\% \citep{muchovej2007}.  Given the large
scatter in the flux density ratios, we adopt a conservative 20\%
calibration uncertainty to account for potential calibration
systematics.

%%%%%%%%%%%%%%%%%%%%%%%%%%%%%%%%%%%%%%%%%%%%%%%%%%%%%%%%%%%%%%%%%%%%%%
\subsection{\chandra\ Observations of \cthreeshort}
\label{subsec:chandra_obs}
%%%%%%%%%%%%%%%%%%%%%%%%%%%%%%%%%%%%%%%%%%%%%%%%%%%%%%%%%%%%%%%%%%%%%%

%\dataset[ADS/Sa.CXO#obs/03248]{Chandra ObsId 3248}
%\dataset[ADS/Sa.CXO#obs/11728]{Chandra ObsId 11728}

\cthreeshortACT\ is the Abell cluster \cthreeshort, which has publicly
available \chandra\ X-ray observations.  The data consist of two
ACIS-I observations, obsIDs \dataset[ADS/Sa.CXO#obs/03248]{3248} and
\dataset[ADS/Sa.CXO#obs/11728]{11728}, of durations 9 and 17~ks
respectively.  The data are reduced starting with the level 1 events
file using CIAO~4.3 and calibration database version~4.4.1.  Standard
corrections are applied, along with light curve filtering and other
standard processing \citep[for reduction details see][]{reese2010}.
The \chandra\ images of \cthreeshort\ are made by binning the
0.7--7.0~keV data to $1.\!\!\arcsec968$ pixels and exposure maps are
computed at 1~keV.  The images and exposure maps from the two
observations are combined, and a wavelet based source detector is used
on the combined image and exposure map to identify and generate a list
of potential point sources.  The list is used as the basis of our
point source mask.

Figure~\ref{fig:a2631_x-sze} shows the smoothed \chandra\ data (color)
along with the deconvolved SZA data (contours).  The color scale shows
\chandra\ counts and the contours are multiples of twice the rms of
the SZA map.  The FWHM of the synthesized beam (effective resolution)
is shown in the lower left corner.  The same E-W elongation seen in
the SZA data is seen in the \chandra\ data as well, and the peaks in
the SZE and X-ray images align.

%%%%%%%%%%%%%%%%%%%%%%%%%%%%%%%%%%%%%%%%%%%%%%%%%%%%%%%%%%%%%%%%%%%%%%%
\section{Analysis}
\label{sec:analysis}
%%%%%%%%%%%%%%%%%%%%%%%%%%%%%%%%%%%%%%%%%%%%%%%%%%%%%%%%%%%%%%%%%%%%%%%

%%%%%%%%%%%%%%%%%%%%%%%%%%%%%%%%%%%%%%%%%%%%%%%%%%%%%%%%%%%%%%%%%%%%%%%
\subsection{Method}
\label{subsec:method}
%%%%%%%%%%%%%%%%%%%%%%%%%%%%%%%%%%%%%%%%%%%%%%%%%%%%%%%%%%%%%%%%%%%%%%%

We fit cluster models to the data and derive parameters from the best-fit
models.  \coneshort\ has no X-ray data and we fit models to the SZE
data only.  \cthreeshort\ has both SZA and \chandra\ data and we
perform three complementary analyses using 1) SZA data only; 2)
\chandra\ data only; and 3) both the SZA and \chandra\ data jointly.
Derived quantities such as mass and integrated SZE signal are computed
within $R_{500}$, the radius within which the mean interior density is
500 times the critical density at the redshift of the cluster,
$\rho_c(z) = 3 H^2(z) / (8\pi G)$, where $H(z)$ is the Hubble
parameter, and $G$ is Newton's gravitational constant.

\subsubsection{\chandra\ Spectroscopy for \cthreeshort}
\label{subsubsec:xspec}

Both a single temperature and a temperature profile are measured for
\cthreeshort.  We extract spectra and response files for both data
sets separately.  Single temperature spectra are extracted within
(0.15--1)$R_{500}$ for the isothermal temperature analyses.  The
region used for spectroscopic extraction is found recursively, by
picking a trial radius, extracting a spectrum to determine the
electron temperature, $T_e$, within that radius, and computing
$M_{500} \equiv M(R_{500})$ from hydrostatic equilibrium
(Eq.~\ref{eq:hse_mass}) and the resulting $R_{500}$.  This process is
repeated until the values of the input and output $R_{500}$ agree to
1\%.  Spectra are also extracted from within the full $R_{500}$
determined as outlined above for comparison.

For the temperature profile, we extract spectra in radial annuli with
$r_{\rm out}/r_{\rm in}$ set to a constant ratio.  This follows
\citet[][hereafter V06]{vikhlinin2006b} who found this choice produces
roughly equal counts per annulus for cluster observations. However, we
choose $r_{\rm out}/r_{\rm in} = 2$ (instead of 1.5 used by V06) in
order to increase the counts in each radial bin, since our data have
many fewer counts than the data considered in
V06. \nocite{vikhlinin2006b}

We construct quiescent ACIS background datasets by reprojecting the
ACIS blank-field observations to match each dataset.  To account for
variations in the particle background, the blank-field observations
are scaled by the ratio of fluxes in the 9.5--12~keV band, where the
\chandra\ effective area is essentially zero and the flux is due
entirely to the background \citep[e.g.,][]{vikhlinin2005}.  Background
spectra are extracted from these quiescent background data sets using
the same regions as that for the cluster data and used when fitting
the cluster spectral properties.

Spectra for both \chandra\ observations using their respective
responses are fit simultaneously to the same plasma model with the
normalizations allowed to vary independently between data sets.  Xspec
\citep{arnaud1996, dorman2001} is used to model the intracluster
medium with an APEC spectrum \citep{smith2001} that includes
Bremsstrahlung and line emission components.  We adopt the Galactic
column density of $N_{\mbox{\tiny H}}=3.55\times 10^{20}$ cm$^{-2}$
\citep{kalberla2005}, solar abundances of \citet{asplund2009}, and
cross sections of \citet{balucinska1992} with an updated He cross
section \citep{yan1998}.  The analysis uses data in the 0.7--7.0~keV
energy range.  The ``cstat'' statistic, which is similar to the
\citet{cash1979} statistic, is used when modeling the spectral data to
properly account for low counts.

Table~\ref{tab:xspec} summarizes the \chandra\ spectral results,
listing the inner and outer extraction radii, derived temperatures,
$T_e$, and metallicity relative to solar, $Z$.  In all cases, the
model provides a good fit to the data, without any obvious structure
or pattern in the residuals.  As a check, the $T_e$ profile spectra
are fit fixing the metallicity to the global value ($Z=0.4$).  The
resulting $T_e$'s agree with those in Table~\ref{tab:xspec} to within
$1\sigma$ for all regions.

%%%%%
%
% X-ray Spectral Properties
%
%%%%%

%% 

\begin{deluxetable}{lccc}

%\singlespace
%\footnotesize
%\rotate
\tablewidth{0pt}
%\tablenum{}
\tablecolumns{4}
%\tableheadfrac{}
\tablecaption{\cthreeshort\ \chandra\ Spectral Properties\label{tab:xspec}}
\tablehead{
\colhead{$R_{\mbox{in}}$} &
\colhead{$R_{\mbox{out}}$} &
\colhead{$T_e$} &
\colhead{$Z$} 
\\
\colhead{(\arcsec)} &
\colhead{(\arcsec)} &
\colhead{(keV)} &
\colhead{($Z_\odot$)} 
}
\startdata
0  & 309 & $7.9 ^{+0.5} _{-0.4}$ & $0.38 ^{+0.14} _{-0.14}$ \\[0.25pc]
\hline
\\[-0.5pc]
46 & 309 & $7.3 ^{+0.6} _{-0.5}$ & $0.42 ^{+0.13} _{-0.12}$ \\[0.25pc]
\hline
\\[-0.5pc]
0   &  20 & $8.3 ^{+2.2} _{-1.5}$ & $1.12 ^{+0.54} _{-0.55}$ \\[0.25pc]
20  &  39 & $8.1 ^{+1.3} _{-1.1}$ & $0.38 ^{+0.31} _{-0.26}$\\[0.25pc]
39  &  79 & $9.9 ^{+1.2} _{-1.1}$ & $0.49 ^{+0.22} _{-0.21}$\\[0.25pc]
79  & 157 & $7.8 ^{+0.8} _{-0.5}$ & $0.48 ^{+0.22} _{-0.17}$\\[0.25pc]
157 & 315 & $6.7 ^{+0.6} _{-1.2}$ & $1.22 ^{+0.36} _{-0.36}$\\[0.25pc]
315 & 630 & $6.0 ^{+1.5} _{-2.2}$ & $0.88 ^{+1.64} _{-0.59}$
\enddata
%\tablenotetext{a}{}
\tablecomments{Fits performed with redshift $z=0.27$
  \citep{struble1999} and Galactic \ion{H}{1} column density
  $N_{\mbox{\tiny H}} = 3.55\times 10^{20}$ cm$^{-2}$
  \citep{kalberla2005}.}
\end{deluxetable}

\subsubsection{Cluster Models}
\label{subsubsec:cluster_model}

For the SZA data, the clusters are fit with both the traditional
$\beta$-model \citep{cavaliere1976, cavaliere1978} and the universal
pressure profile of \citet[][hereafter A10]{arnaud2010}.  For the
SZE-only $\beta$-model analysis, we fix $\beta=0.86$ from a best fit
to an average SZE profile \citep{plagge2010}.

For the \chandra\ data, the cluster is fit with the traditional
isothermal $\beta$-model and the modified $\beta$-model (e.g., V06)
\nocite{vikhlinin2006b} for density along with the corresponding V06
$T_e(r)$ profile, hereafter referred to as the V06 model.  Many of the
V06 model parameters must be fixed, given the large number of
parameters compared with the quality of the X-ray data.
\cthreeshort\ does not show signs of a cool core so we excise those
parts of the V06 model.

Joint fits to the SZA and \chandra\ data are performed by assuming the
A10 pressure profile and a simplified, core-cut form of the V06
density profile (hereafter sV06) for the cluster models.  These fits
are performed both including and excluding the X-ray spectroscopic
data, allowing the SZE and X-ray surface brightness data to place
constraints on the cluster temperature \citep[for details
  see][]{mroczkowski2009}.  The temperature is derived assuming the
ideal gas law ($P_e = n_e k_B T_e$).  The X-ray surface brightness
data determine the sV06 density fit, and the SZE surface brightness
data drive the A10 pressure profile fit.  A single temperature is
derived from the inferred $T_e$ profile using the prescription of
\citet{mazzotta2004} over the (0.15-1)$R_{500}$ region.  For the
A10+sV06 fits that include spectroscopic information, the likelihood
from the inferred $T_e$ described above is included in the MCMC using
the output from the \chandra\ single-temperature spectroscopy.  A
separate MCMC run is done without including the $T_e$ spectral
component in order to assess how X-ray surface brightness and SZE data
can be used to find $M(r)$ without relying on X-ray spectroscopy.

\subsubsection{Radio Point Source Model}
\label{subsubsec:pt_src_model}

The radio point source model accounts for the primary beam
attenuation and includes a spectral index that models the
frequency dependence.  The spectral dependence of the point source
model is given by
\begin{equation}
  F_\nu=F_{\nu_0} \left ( \frac{\nu}{\nu_0} \right ) ^{\alpha},
  \label{eq:pt_flux_mod}
\end{equation}
where $F_\nu$ is the intrinsic point source flux density at frequency
$\nu$, $F_{\nu_0}$ is the intrinsic flux density at fiducial frequency
$\nu_0$, and $\alpha$ is the spectral index.  We adopt the average
frequency for the fiducial frequency, $\nu_0 = 30.938$~GHz, and report
flux densities at this frequency.

Point sources are first modeled with both $F_{\nu_0}$ and $\alpha$
allowed to vary.  The 8~GHz bandwidth of the SZA potentially provides
leverage on both the spectral index and flux density of each point
source.  The flux density is, in general, well constrained.  However,
only weak constraints are obtained on $\alpha$ for all but the
brightest sources.  We first run fits including $\alpha$ in the Markov
chain and then determine $\alpha$ using our 30~GHz flux density
combined with 1.4~GHz flux densities from the NRAO VLA Sky Survey
\citep[NVSS;][]{condon1998}.  As a check, we compare the flux
densities obtained from the NVSS-based spectral indices to those
obtained just from the SZA data.  The flux densities are consistent
within the 68\% confidence regions.

\subsubsection{Markov Chain Monte Carlo Analysis}
\label{subsubsec:mcmc}

We perform a Markov chain Monte Carlo (MCMC) analysis of the SZE and
X-ray data \citep[for details see][]{reese2000, bonamente2004}.  The
philosophy behind the analysis is to keep the data in a reduced but
native state, and to run the models through the observing strategy to
compare directly to the data.  Interferometers measure the Fourier
transform of the sky brightness modulated by the primary beam.
Therefore the SZA data most naturally provide constraints in the
$u$-$v$ plane, where the noise properties of the data and the spatial
filtering of the interferometer are well understood.  We perform model
fits directly in the $u$-$v$ plane, computing the Gaussian SZE
likelihood, $\mathcal{L}_{\mathrm{SZE}}$.  For X-ray data, the Poisson
likelihood, $\mathcal{L}_{\mathrm{X}}$, is computed for each pixel,
ignoring those that fall within the point source mask.  The SZE and
X-ray data are independent, and their likelihoods can simply be
multiplied to obtain the combined likelihood.  Best-fit parameters and
confidence intervals are determined from the 50\%, 16\%, and 84\%
levels of the cumulative distribution function, which define the the
median and 68\% confidence region.  The resultant probability
distributions are visually inspected, and convergence and mixing of
the Markov chains are checked with the Geweke Z-statistic
\citep{geweke1992}.

%%%%%
%
% Point sources
%
%%%%%

%% 
%%
%% Field Obs  RA   DEC   F30   F1.4  \alpha  F148(est)
%% 
%% 
%%
%%

\begin{deluxetable*}{lccccccc}

%\singlespace
%\footnotesize
%\rotate
\tablewidth{0pt}
%\tablenum{}
\tablecolumns{7}
%\tableheadfrac{}
\tablecaption{Radio Point Sources\label{tab:pt}}
\tablehead{
\colhead{} &
\colhead{R.A. (J2000)} &
\colhead{Decl. (J2000)} &
\colhead{$R$\tablenotemark{a}} &
\colhead{$F_{\mbox{\tiny\rm 31}}$} &
\colhead{$F_{\mbox{\tiny\rm 1.4}}$} &
\colhead{} &
\colhead{$F_{\mbox{\tiny\rm 148}}^{est}$}
\\
\colhead{Field} &
\colhead{(h m s)} &
\colhead{(d m s)} &
\colhead{($\arcmin$)} &
\colhead{(mJy)} &
\colhead{(mJy)} &
\colhead{$\alpha$} &
\colhead{(mJy)}
}
\startdata
\coneshort 
        & 00 22 13.006
	& -00 36 33.35 
        & 0.3
	& $1.33 ^{+0.15} _{-0.15}$
	& $18.8 ^{+0.7} _{-0.7}$
        & $-0.9 \pm 0.1$ 
        & $0.3$
\\[.25pc]
        & 00 22 05.987
        & -00 35 54.95  
        & 1.6
        & $0.98 ^{+0.13} _{-0.13}$
        & $33.2 ^{+1.1} _{-1.1}$
        & $-1.1 \pm 0.1$ 
        & $0.2$
\\[.25pc]
\ctwoshort 
        & 20 51 17.758
	& +00 53 20.16
        & 2.3
	& $7.36 ^{+0.14} _{-0.14}$
	& $43.8 ^{+1.4} _{-1.4}$
        & $-0.6 \pm 0.1$ 
        & $3.0$
\\[.25pc]
        & 20 51 40.720
	& +00 52 02.74 
        & 7.3
	& $4.01 ^{+0.44} _{-0.44}$
	& $16.9 ^{+0.7} _{-0.7}$
        & $-0.5 \pm 0.1$ 
        & $1.9$
\\[.25pc]
\cthreeshort 
        & 23 37 40.209
	& +00 16 42.18
        & 0.7
        & $3.99 ^{+0.16} _{-0.17}$
	& $107.0 ^{+3.8} _{-3.8}$
        & $-1.1 \pm 0.1$ 
        & $0.8$
\\[.25pc]
        & 23 37 38.106
	& +00 10 01.93
        & 6.2
        & $3.20 ^{+0.34} _{-0.33}$
	& $12.4 ^{+0.6} _{-0.6}$
        & $-0.4 \pm 0.1$ 
        & $1.6$
\enddata
\tablenotetext{a}{Projected distance from the cluster center.}
%\tablecomment{}
\end{deluxetable*}

The cluster and any detected point sources in the field are modeled
for the SZA data.  The SZE signal varies as
\begin{equation}
\Delta T_{\mbox{\tiny SZE}} = f(\nu) T_{\mbox{\tiny CMB}} \int d\ell 
\frac{k_{\mbox{\tiny B}}T_e}{m_e c^2} n_e \sigma_{\mbox{\tiny T}} \equiv f(\nu)
T_{\mbox{\tiny CMB}} y,
\label{eq:sze_defined}
\end{equation}
where $f(\nu)$ is the frequency dependence of the SZE at frequency
$\nu$, $n_e$ and $T_e$ are the electron number density and
temperature, $k_{\mbox{\tiny B}}$ is the Boltzmann constant, $m_e$ is
the mass of the electron, $\sigma_{\mbox{\tiny T}}$ is the Thomson
cross section, integration is along the line of sight $\ell$, and $y$
is the Compton-$y$ parameter.  SZA data have sixteen 500 MHz bands
covering 8~GHz of bandwidth.  The frequency dependence of the SZE is
taken into account when modeling.  Relativistic corrections
to $f(\nu)$ \citep[e.g.,][]{itoh1998, challinor1998} depend on $T_e$
and are not included in this analysis for consistency, since only one
cluster has a measured $T_e$.  The effects of this small ($\lesssim
3$\% at 30~GHz) correction are discussed in
Section~\ref{sec:discussion}.

For \cthreeshort, \chandra\ data are modeled with a cluster model and
a constant X-ray background.  Regions containing point sources are
excluded from the fit.

\subsubsection{Derived Cluster Properties}
\label{subsubsec:derived_prop}

Derived quantities such as the integrated Compton-$y$, $Y_{500}$, and
mass, $M_{500}$, are computed for each step in the Markov chain for
each of the above types of fits we perform.  We use these output
chains to determine the median values and 68\% confidence regions for
each parameter of interest.  This method cleanly propagates
uncertainties from the parameters included in the chain.  For example,
uncertainties from modeling detected radio sources are propagated
into the integrated Compton-$y$ results.

We compute the integrated Compton-$y$, $Y_{\mbox{\tiny sph}}(r) =
{\sigma_{\mbox{\tiny T}}}/{(m_{\rm e} c^2)}\int\! P_e dV$, within a
spherical volume enclosed by radius $r$, where $dV = 4 \pi r^2 dr$ and
$P_e \equiv n_e k_{\mbox{\tiny B}} T_e$ is the electron pressure.
Since this latter quantity tracks thermal energy ($E=3/2 \int \!
P_{\mbox{\tiny gas}} dV$), and thermal pressure is the dominant source of
support against gravitational collapse \cite[see, e.g.][who report
  that only 10-20\% of the pressure comes from non-thermal
  support]{nagai2007, lau2009}, we can expect $Y_{\mbox{\tiny
    sph}}(r)$ to track gravitational energy within radius $r$.
Assuming the virial relation, a constant gas fraction of 0.13, an
average metallicity, and a total mass profile that can be described
by a Navaro, Frenk, and White (NFW) halo model \citep{navarro1996,
  navarro1997}, we estimate mass from fits to $Y_{\mbox{\tiny
    sph}}(r)$ derived from our A10 and $\beta$-model fits to the SZE
data \citep[as was done in][]{mroczkowski2011}.  These SZE-only mass
estimates can be performed regardless of chosen model fit, and, like
estimates from the X-ray assuming hydrostatic equilibrium, rely on
spherical symmetry and do not take into account sources of non-thermal
pressure support.  We define $Y_{500} \equiv Y_{\mbox{\tiny
    sph}}(R_{500})$ as the integrated Compton-$y$ within $R_{500}$.

Mass estimates from the \chandra\ data are based on hydrostatic
equilibrium \citep[e.g.,][]{sarazin1988},
\begin{equation}
 M(r) = -\frac{k_{\mbox{\tiny B}} T_e(r) r}{G\mu m_p} \left (
 \frac{d\ln(n_e)}{d\ln(r)} + \frac{d\ln(T_e)}{d\ln(r)}\right ) = 
-\frac{r^2}{G \rho_{\mbox{\tiny gas}}(r)} \frac{dP_{\mbox{\tiny gas}}(r)}{dr},
\label{eq:hse_mass}
\end{equation}
where $\mu$ is the mean molecular weight, $m_p$ is the mass of the
proton, and $\rho_{\mbox{\tiny gas}}$ and $P_{\mbox{\tiny gas}}$ are
the total gas density and pressure, respectively.  The mass as a
function of radius is used to compute $R_{500}$ using $M_{500} \equiv
M(R_{500}) = (4\pi/3) R_{500}^3 500 \rho_c(z) $, where $\rho_c(z)$ is
the critical density at redshift $z$.  Our analysis is similar to that
of \citet{vikhlinin2006b}.

%%%%%%%%%%%%%%%%%%%%%%%%%%%%%%%%%%%%%%%%%%%%%%%%%%%%%%%%%%%%%%%%%%%%%%%
\subsection{Results}
\label{subsec:results}
%%%%%%%%%%%%%%%%%%%%%%%%%%%%%%%%%%%%%%%%%%%%%%%%%%%%%%%%%%%%%%%%%%%%%%%

%%%%%
%
% Mass estimates:  SZE & X-ray
%
%%%%%

%% 
%%
%%
%% 
%% 
%%
%%

\begin{deluxetable}{lccc}

%\singlespace
%\footnotesize
%\rotate
\tablewidth{0pt}
%\tablenum{}
\tablecolumns{4}
%\tableheadfrac{}
\tablecaption{SZE and X-ray Derived Properties\label{tab:mass}}
\tablehead{
\colhead{Field} &
\colhead{$R_{500}$} &
\colhead{$M_{500}$} &
\colhead{$Y_{500}$}
\\
\colhead{(model)} &
\colhead{(Mpc)} &
\colhead{($10^{14} \rm M_\odot$)} &
\colhead{($10^{-4}$ Mpc$^2$)} 
}
\startdata
\bf \coneshort \\ 
~~$\beta$--SZE & $1.07 ^{+0.04}_{-0.04}$ & $8.8 ^{+1.1}_{-1.0}$ & $1.04^{+0.20}_{-0.16}$\\[.25pc]
~~A10--SZE   & $1.00 ^{+0.05}_{-0.05}$ & $7.3 ^{+1.0}_{-1.0}$ & $0.84^{+0.16}_{-0.14}$
\\
\bf \cthreeshort \\
~~$\beta$--SZE &
 $1.42 ^{+0.15} _{-0.13}$ & $11.5 ^{+4.1} _{-2.8}$ & $1.38 ^{+0.85} _{-0.49}$\\[.25pc]
~~A10--SZE &
 $1.58 ^{+0.36} _{-0.25}$ & $16.0 ^{+13.5}_{-6.5}$ & $1.58 ^{+1.41} _{-0.65}$\\[.25pc]
~~$\beta$--X-ray &
 $1.31 ^{+0.03} _{-0.03}$ & ~$8.5 ^{+0.8} _{-0.7}$ & \nodata\\[.25pc]
~~V06--X-ray &
 $1.54 ^{+0.10} _{-0.11}$ & $13.8 ^{+2.9} _{-2.7}$ & \nodata\\[.25pc]
~~A10+sV06\tablenotemark{a} &
 $1.33 ^{+0.20} _{-0.12}$ & ~$9.4 ^{+4.8} _{-2.4}$ & $1.03 ^{+0.31} _{-0.17} $\\[.25pc]
~~A10+sV06\tablenotemark{b} &
 $1.31 ^{+0.35} _{-0.38}$ & ~$9.1 ^{+9.4} _{-5.9}$ & $1.01 ^{+0.65} _{-0.52}$\\[-.6pc]
\enddata
%\tablecomment{}
\tablenotetext{a}{Includes X-ray spectroscopic information.}
\tablenotetext{b}{Does not include X-ray spectroscopic information.}
\end{deluxetable}

%%%%%
% SZA u-v radial profile
%%%%%

%%%%%%%%%%%%%%%%%%%%%%%%%%%%%%%%%%%%%%%%%%%%%%%%%%%%%%%%%%%%%%%%%%%%%%%
\begin{figure*}[!tbh]
  \centerline{
    \includegraphics[width=3.3in]{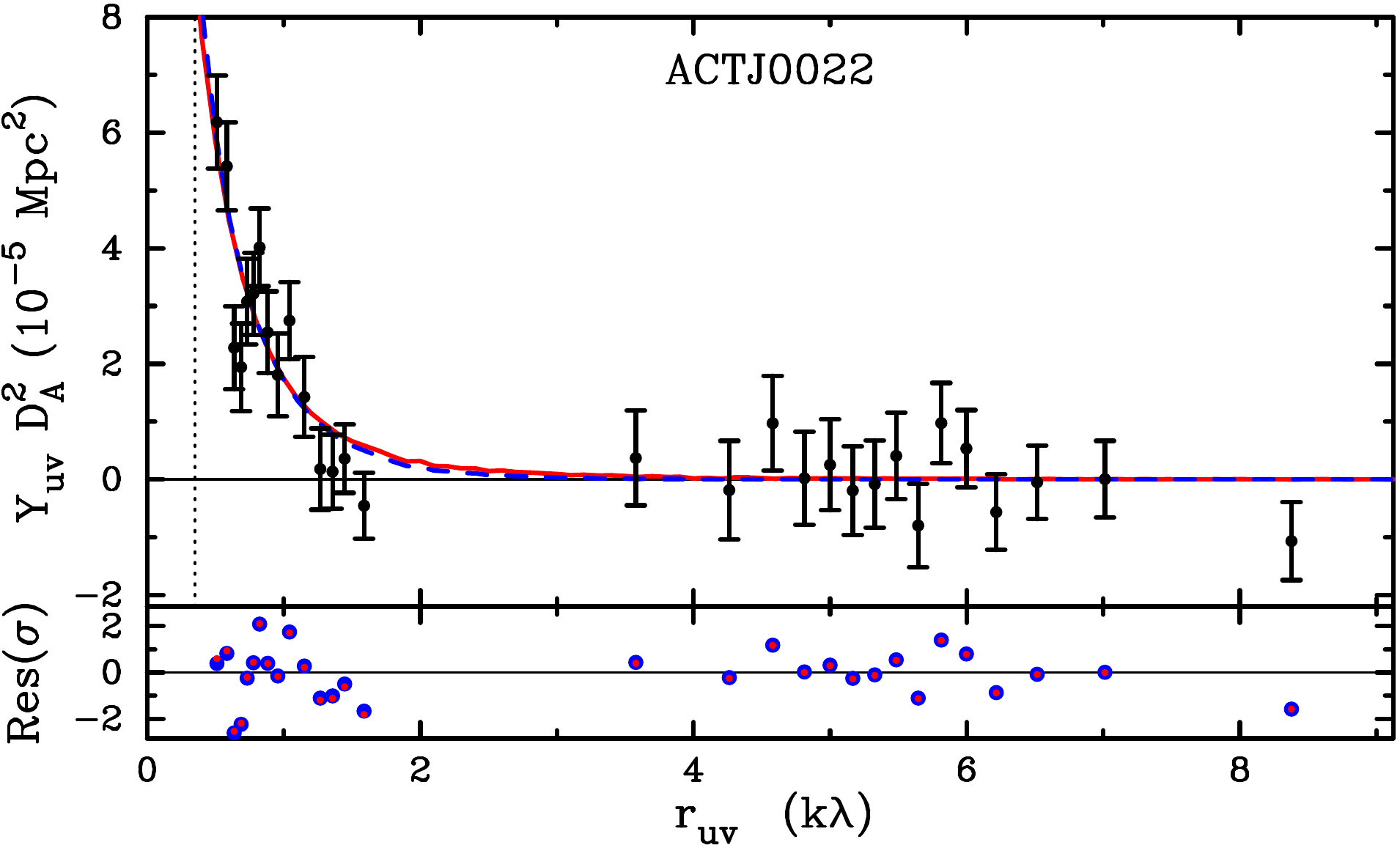}
    \includegraphics[width=3.3in]{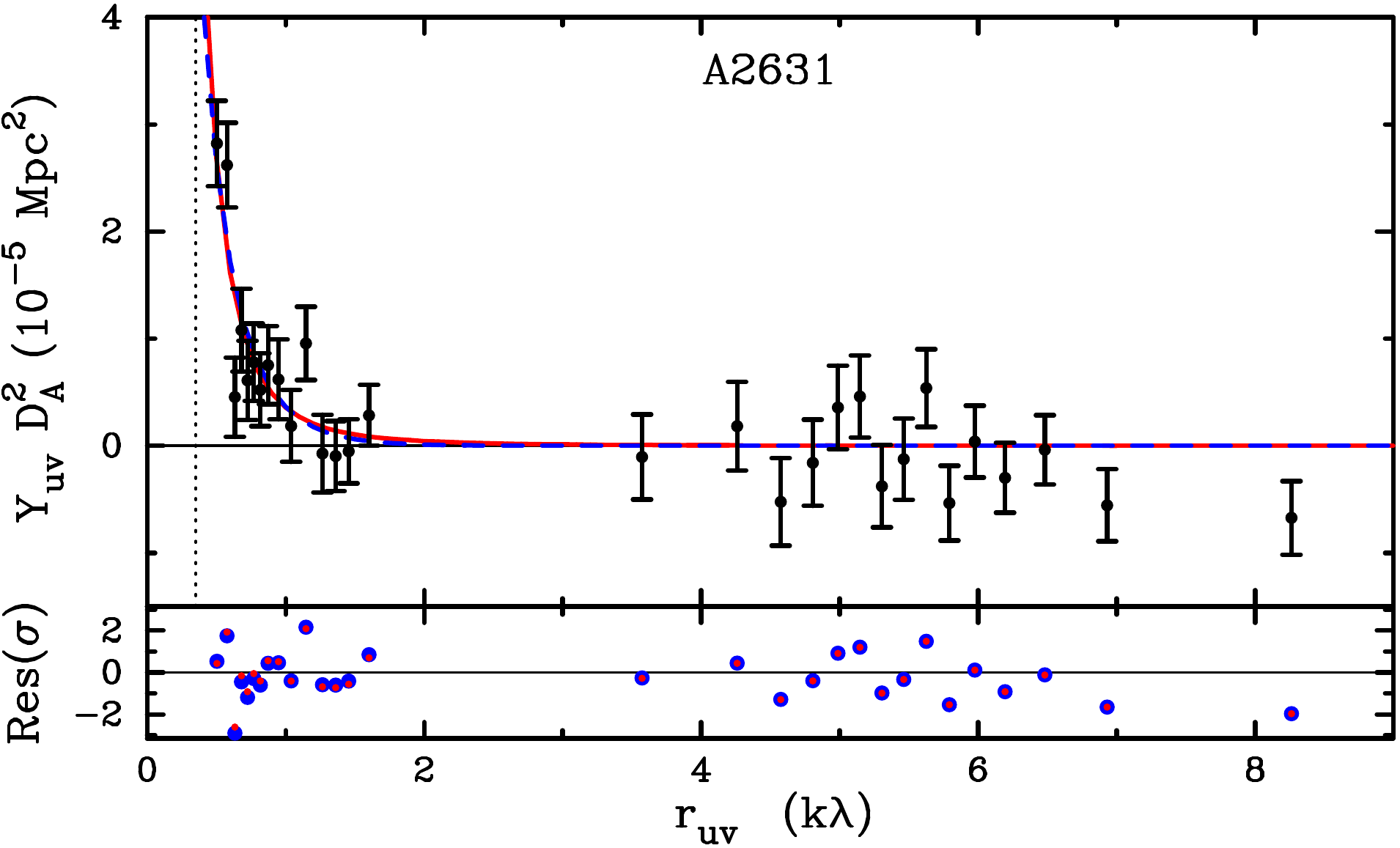}
  }
  \caption{SZE profiles for \coneshort\ (left) and
    \cthreeshort\ (right) as a function of $u$-$v$ radius,
    $r_{uv}=\sqrt{u^2+v^2}$.  Best-fit A10 (red) and $\beta$ (blue)
    model fits are also shown. Residuals are shown in the lower
    sections in units of the standard deviation.  Point sources have
    been subtracted from the visibilities and the phase centers
    shifted to the cluster centers before computing the profiles.  The
    $u$-$v$ plane Compton-$y$, $Y_{uv}$, is scaled by $D_{\mbox{\tiny
        A}}^2$, creating a SZE luminosity-like quantity \citep[for
      details see, e.g.,][]{mroczkowski2009}.  The real parts of
    $Y_{uv}$ are shown.  The imaginary parts are consistent with zero.
    The dotted line shows the shadowing limit of the 3.5~m dishes.
    \label{fig:a2631_szeprof}}
\end{figure*}
%%%%%%%%%%%%%%%%%%%%%%%%%%%%%%%%%%%%%%%%%%%%%%%%%%%%%%%%%%%%%%%%%%%%%%%

%%%%%
% Chandra surface brightness profile
%%%%%

%%%%%%%%%%%%%%%%%%%%%%%%%%%%%%%%%%%%%%%%%%%%%%%%%%%%%%%%%%%%%%%%%%%%%%%
\begin{figure}[!tbh]
  \centerline{
    \includegraphics[width=3.5in]{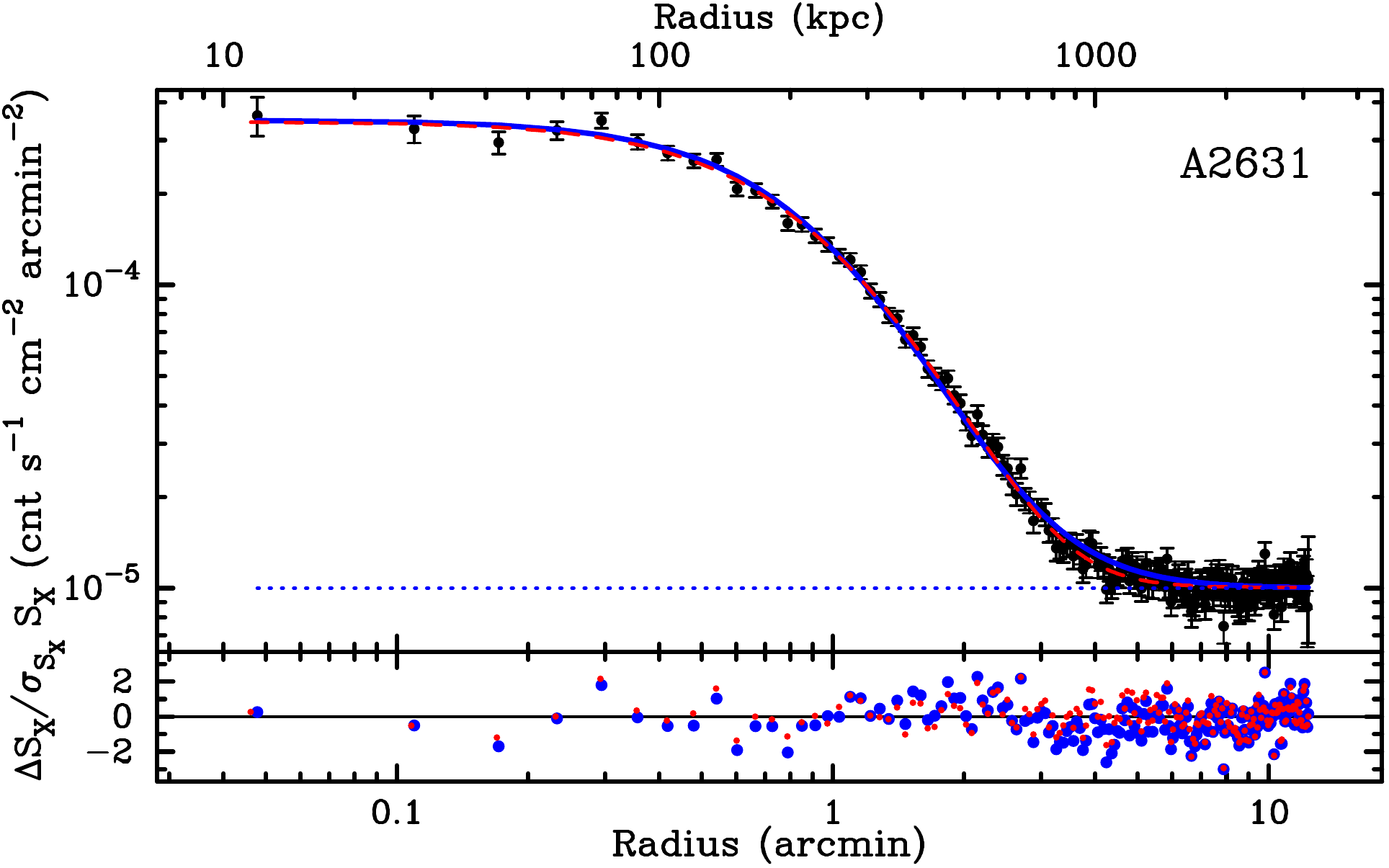}
  }
  \centerline{
    \includegraphics[width=3.5in]{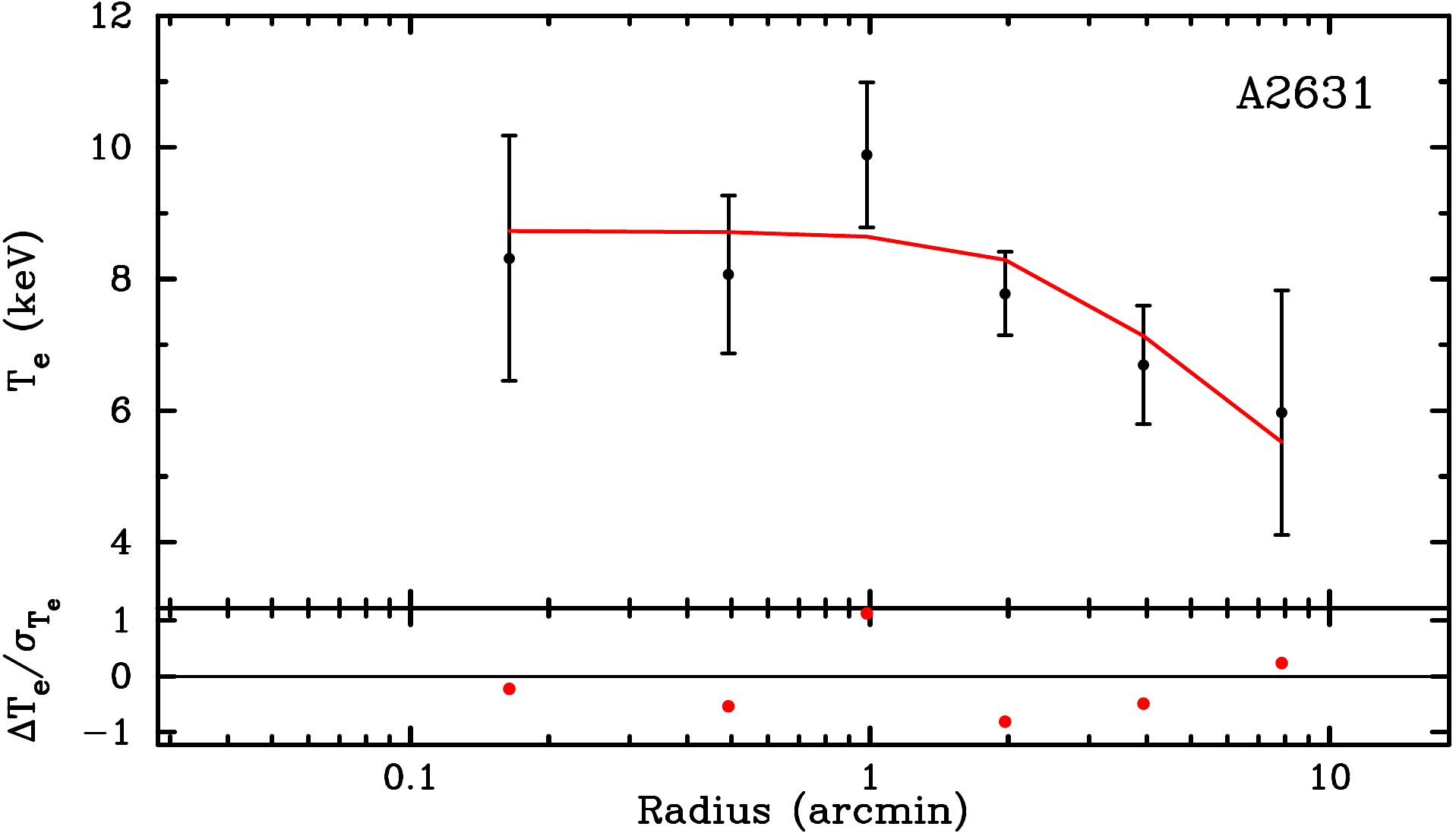}
  }
  \caption{{\it Upper:} \cthreeshort\ \chandra\ surface brightness
    profile (points) with best fit $\beta$ (blue) and V06 (red)
    models.  The lower section shows residuals in units of the
    standard deviation.  The models are similar and the simple
    spherical $\beta$-model provides a good fit to the \chandra\ data.
    {\it Lower:} \cthreeshort\ \chandra\ temperature profile (points)
    with best fit V06 temperature (red) model.  The lower section
    shows residuals in units of the standard deviation.
    \label{fig:a2631_xprof}}
\end{figure}
%%%%%%%%%%%%%%%%%%%%%%%%%%%%%%%%%%%%%%%%%%%%%%%%%%%%%%%%%%%%%%%%%%%%%%%

The MCMC results for all detected point sources are presented in
Table~\ref{tab:pt}.  In addition to the SZA 31~GHz flux densities, the
corresponding NVSS \citep{condon1998} 1.4~GHz flux densities are
listed.  These two flux densities are used to compute the spectral
index $\alpha$, which is then used to estimate the point source flux
density in ACT's 148~GHz band, $F_{\mbox{\tiny 148}}^{est}$ (both also
listed in Table~\ref{tab:pt}).  The projected distance from the
cluster, $R$, is also listed and gives an idea of the potential impact
of each source on cluster detection and potential contamination of the
SZE flux.  Table~\ref{tab:mass} summarizes the cluster modeling
results for the SZA, \chandra, and joint analyses.  We report
$R_{500}$, $M_{500}$, and $Y_{500}$ from our MCMC analysis.

Figure~\ref{fig:a2631_szeprof} shows the SZE $u$-$v$ radial profiles
of the SZA data (points) with best-fit $\beta$ (blue) and A10 (red)
models for \coneshort\ (left) and \cthreeshort\ (right).  The
visibilities ($u$-$v$ plane data) are converted to a frequency
independent $u$-$v$ plane Compton-$y$, $Y_{uv}$, and scaled by the
angular diameter distance squared, $D_{\mbox{\tiny A}}^2$, creating a
SZE luminosity-like quantity \citep[for details see,
  e.g.,][]{mroczkowski2009}.  The real parts of $Y_{uv}$ are plotted.
The imaginary components are consistent with zero. Residuals are shown
in the lower sections of both panels.  The \coneshort\ radial profile
has $\chi^2=35.1$ and 35.8 with 28 degrees of freedom (dof) for the
$\beta$ and A10 models, respectively.  The corresponding probabilities
of obtaining these $\chi^2$'s or larger by chance given the degrees of
freedom are, $p(\geq \chi^2 | \, \mbox{dof})=0.17$ and $0.15$.  The
\cthreeshort\ radial profiles have $\chi^2=38.1$ and 36.1 with 28 dof
for the $\beta$ and A10 models, respectively, with corresponding
$p(\geq \chi^2| \, \mbox{dof})=0.10$ and $0.14$.

Figure~\ref{fig:a2631_xprof} shows the \chandra\ surface brightness
profile (upper) and temperature profile (lower) along with the best
fit models.  The data (points) and best fit $\beta$ (blue) and V06
(red) models are shown.  The best-fit background level is shown by the
dotted line.  The lower portion of each panel shows the residuals in
units of the standard deviation.  Despite this cluster's E-W
elongation (Fig.~\ref{fig:a2631_x-sze}), a spherical model, on
average, provides a good fit with $\chi^2=190.3$ and $155.9$ for the
$\beta$ and V06 radial profiles that have 194 and 191 degrees of
freedom (dof), giving $p(\geq \chi^2 | \, \mbox{dof})=0.56$ and
$0.97$, respectively.  The V06 temperature profile results in
$\chi^2=2.6$ for 2 degrees of freedom with $p(\geq \chi^2 | \,
\mbox{dof})=0.27$.

%%%%%%%%%%%%%%%%%%%%%%%%%%%%%%%%%%%%%%%%%%%%%%%%%%%%%%%%%%%%%%%%%%%%%%%
\subsection{Optically Informed Cluster Properties}
\label{subsec:mass_optical}
%%%%%%%%%%%%%%%%%%%%%%%%%%%%%%%%%%%%%%%%%%%%%%%%%%%%%%%%%%%%%%%%%%%%%%%

SDSS S82 data provide redshift information and enable mass estimates
based on relations between cluster mass and optical properties such as
richness and luminosity \citep[e.g.,][]{johnston2007, reyes2008,
  rozo2009}.  We have computed cluster masses from the
$N_{200\overline{\rho}}$--$M_{500}$ scaling relation of
\citet{rozo2009}, which is based on weak-lensing and X-ray
measurements of clusters from the SDSS maxBCG catalog
\citep{koester2007}.  Masses computed from the
$N_{200\overline{\rho}}$--$M_{500}$ scaling relation of
\citet{reyes2008} yield similar results after correcting for the
different radii used between the studies.  $M_{500}$ is the mass
within $R_{500}$, as defined earlier, the radius within which the mean
interior density is 500 times that of the critical density at the
redshift of the cluster.  $N_{200\overline{\rho}}$ is the number of
red sequence galaxies in a cluster measured within
$R_{200\overline{\rho}}$, the radius within which the mean interior
density is $200$ times the mean matter density at the redshift of the
cluster, $\overline{\rho}(z)=\Omega_M(z) \, \rho_c(z)$, and is denoted
by the subscript $\overline{\rho}$.  $R_{200\overline{\rho}}$ at $z=0$
corresponds to $R_{60}$ (60 with respect to critical), and is
substantially larger than $R_{500}$ and $R_{200}$.  Cluster
$N_{200\overline{\rho}}$'s are computed for the S82 data using the
maxBCG prescription as implemented by \citet[][see Section~2.2 for
  details]{menanteau2010b}.  Table~\ref{tab:optical_properties}
summarizes the measured $N_{200\overline{\rho}}$ and inferred values
of $M_{500}$ and $R_{500}$ for each cluster field.

%%%%%
%
% Optical properties
%
%%%%%

%% 
%%
%%  Field  N200  M500 R500  Y500 (X500's from Rozo et al.. 2009)
%% 
%% 
%%
%%

\begin{deluxetable}{lcccc}

%\singlespace
%\footnotesize
%\rotate
\tablewidth{0pt}
%\tablenum{}
\tablecolumns{4}
%\tableheadfrac{}
\tablecaption{Optically Informed Properties \label{tab:optical_properties}}
\tablehead{
\colhead{} &
\colhead{} &
\colhead{$M_{500}$} &
\colhead{$R_{500}$} &
\colhead{$Y_{500}$}
\\
\colhead{Field} &
%\colhead{$N_{200}$} &
\colhead{$N_{200\overline{\rho}}$} &
\colhead{($\times 10^{14} \rm M_\odot$)} &
\colhead{(Mpc)} &
\colhead{($\times 10^{-4}$ Mpc$^2$)}
}
\startdata
\coneshort %& $112.99 \pm 10.63$ 
& $113 \pm 11$ & $7.77 \pm 1.12$ & $1.03 \pm 0.05$ &
$0.84 ^{+0.11}_{-0.11}$\\
\ctwoshort %&  $24.52 \pm 4.95$ 
&  $25 \pm 5$  & $1.54 \pm 0.33$ & $0.73 \pm 0.05$ & 
\nodata \\
\cthreeshort %&$77.23 \pm 8.79$  
&$77 \pm 9$  & $5.19 \pm 0.75$ & $1.11 \pm 0.05$ & 
$1.43 ^{+1.01} _{-0.56}$\\[-.6pc]
%\\
\enddata
%\tablenotetext{a}{Gaussian taper with FWHM of 2000 $\lambda$ for OVRO
%data and 1000 $\lambda$ for BIMA data.}
\tablecomments{$N_{200\overline{\rho}}$'s
%$N_{200}$'s 
are measured from SDSS S82 data, $M_{500}$ is derived from
$N_{200\overline{\rho}}$ using the scaling relation of
\citet{rozo2009}, and $R_{500}$ is inferred from the mass.  This
$R_{500}$ is then used to compute $Y_{500}$ from the SZA data for
comparison to the Planck results \citep{planck2011h}.}

\end{deluxetable}

\begin{figure}[!tbh]
  \centerline{
    \includegraphics[width=3.0in]{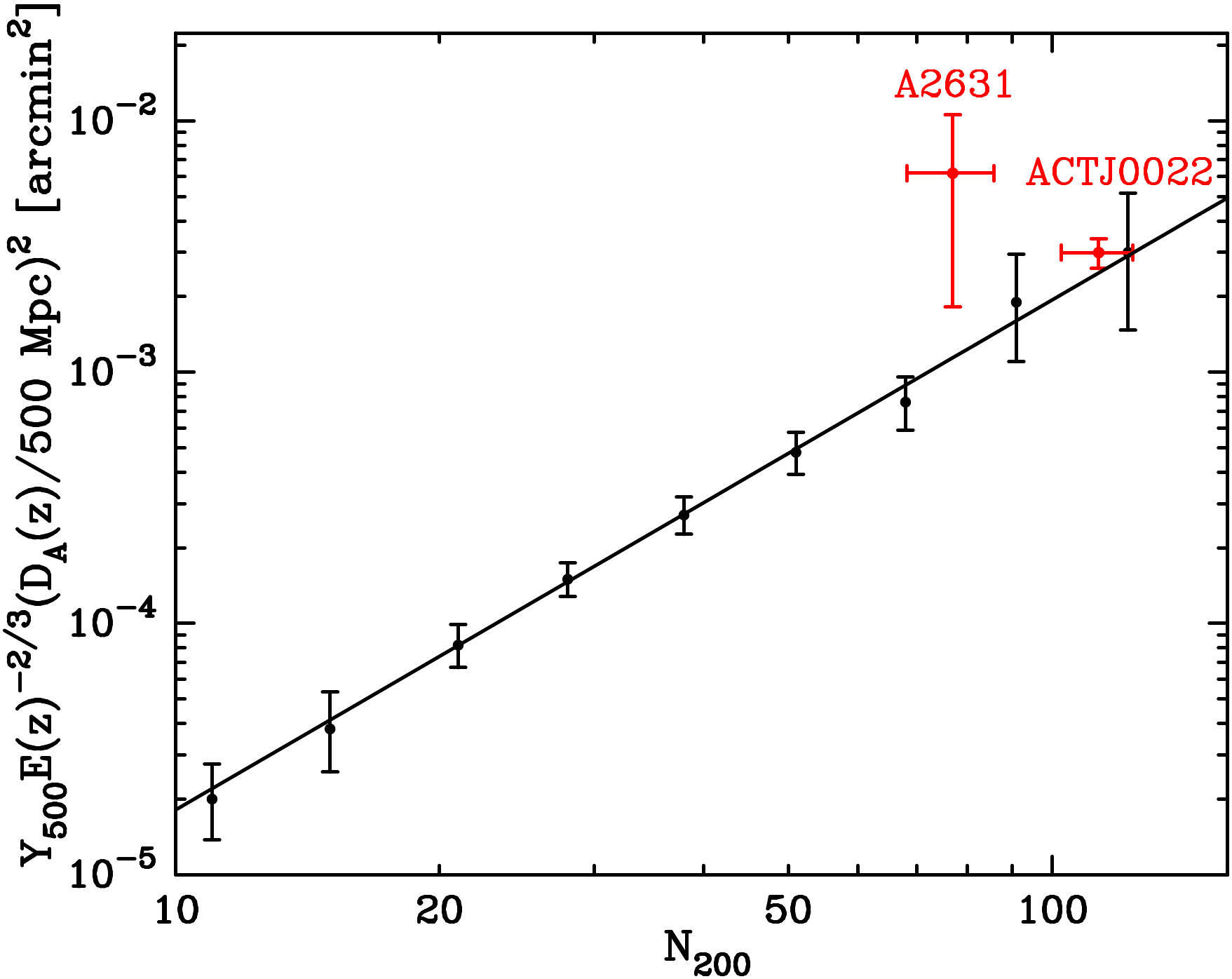}
  }
  \caption{Integrated Compton-$y$--optical richness relation, 
    $\tilde{Y}_{500}$--$N_{200\overline{\rho}}$, for
    Planck (black) and this work (red).  Also shown is the best fit
    relation to the Planck data (solid line).  Integrated
    Compton-$y$ values are spherical values within $R_{500}$ scaled to
    redshift zero and a distance $D_{\mbox{\tiny A}}=500$ Mpc.
    Richness, $N_{200\overline{\rho}}$, is the number of red sequence
    galaxies within $R_{200\overline{\rho}}$.
    \label{fig:y500_n200}}
\end{figure}

A recent \planck\ study explores the SZE-optical scaling relations by
employing a multi-frequency matched filter on \planck\ maps at the
positions of the SDSS maxBCG clusters \citep{planck2011h}.  This work
finds an offset between the measured integrated Compton-$y$--optical
richness relation compared to model predictions for the full maxBCG
sample.  However, when using the X-ray subsample, the authors find good
agreement between the prediction and the model.  Following
\citet{planck2011h}, we determine $R_{500}$ from the optical
properties via the $N_{200\overline{\rho}}$--$M_{500}$ relation
(Table~\ref{tab:optical_properties}) and compute $Y_{500}$ within that
radius, using the fits of the A10 profile to the SZA data.  $Y_{500}$
is rescaled to redshift $z=0$ using the evolution of the Hubble
parameter for a flat universe, $E(z) = \sqrt{\Omega_M(1+z)^3 +
  \Omega_\Lambda}$, and a fiducial distance $D_{\mbox{\tiny A}} = 500$
Mpc, as
\begin{equation} 
\tilde{Y}_{500} \equiv Y_{500} E(z)^{-2/3} (D_{\mbox{\tiny A}}(z)/500
\rm~Mpc)^2\ [arcmin^2].
\label{eq:Ytilde}
\end{equation}
Figure~\ref{fig:y500_n200} compares our two ACT clusters (red points)
to those of \planck\ (black points) for the
$\tilde{Y}_{500}$--$N_{200\overline{\rho}}$ relation.  The best fit
power law for the \planck\ data over the full maxBCG catalog is also
shown (line).  The known X-ray cluster, \cthreeshort, lies above the
\planck\ relation, though with large uncertainty.

%%%%%%%%%%%%%%%%%%%%%%%%%%%%%%%%%%%%%%%%%%%%%%%%%%%%%%%%%%%%%%%%%%%%%%
\section{Discussion}
\label{sec:discussion}
%%%%%%%%%%%%%%%%%%%%%%%%%%%%%%%%%%%%%%%%%%%%%%%%%%%%%%%%%%%%%%%%%%%%%%

The density of compact radio sources is higher in cluster regions
compared to the field \citep[][]{cooray1998a, massardi2004, lin2007,
  coble2007}.  Contamination of the SZE signal from radio sources
could potentially bias flux and mass estimates from the SZE.  The
radio source fluxes extrapolated to 148~GHz (Table~\ref{tab:pt})
provide information on potential contamination of the SZE decrement
signal.  The two sources with projected radius from the cluster
$R>6\arcmin$ will have little impact on cluster detection and flux.
ACT 148~GHz equatorial maps have a noise level of around 2~mJy
beam$^{-1}$ so that most sources are expected to fall well below the
ACT map noise level, and the brightest source is expected to be 1.5
times the noise level. Very Large Array (VLA) observations of galaxies
in nearby clusters between 5 and 40~GHz find that about 60\% of the
point sources show a flattening of the spectral shape above 8~GHz
\citep{lin2009}.  This implies that extrapolating from low frequency
to high frequency yields a lower limit on the contaminating flux.  A
conservative upper limit on the radio source flux density at 148~GHz
is obtained by increasing the extrapolated estimates by a factor of
two.  However, recent simulations of the microwave sky suggest that
only 3\% of clusters have their 148~GHz SZE decrements contaminated at
the $\gtrsim 20$\% level \citep{sehgal2010}.  There is no indication
in the ACT data of contamination by sources in these three cluster
fields.  The estimated 148~GHz flux densities suggest that radio
sources do not significantly impact cluster detection in surveys for
the brightest clusters, although they could potentially bias flux
measurements in some fraction of the clusters at a level of 3--6~mJy
($\lesssim 20$\% for typical $-30$~mJy integrated SZE fluxes).  We
reiterate that extrapolation to higher frequencies is uncertain and
note that this is based on sources in only 3 cluster fields.  More
precise flux density estimates at ACT frequencies will be obtained
through observing a larger number of ACT cluster fields over a range
of frequencies.

The VLA FIRST survey \citep{becker1995} covers the \coneshort\ and
\cthreeshort\ fields.  Three out of the 4 radio point sources have
NVSS and FIRST flux densities that agree within 68\% confidence.  The
107~mJy NVSS source in \cthreeshort's field has a 91~mJy flux density
in the FIRST catalog, resulting in a predicted 148~GHz flux density
that is 9\% higher than that estimated from the NVSS flux density.
These surveys are in different configurations and observe the same
fields at different times, thus providing a rough handle on potential
time-variability of these sources, which could impact the
contamination of the SZE signal and our use of point sources to
calibrate our SZA data (Section~\ref{subsec:sza_cal}).  A 10\% flux
density variability is encompassed by our choice of a 20\% calibration
uncertainty for the SZA data.

The S82 data for \ctwoshort\ exhibit a red sequence typical of galaxy
clusters and provide a spectroscopic redshift.  The
optical mass estimates from the S82 data suggest that \ctwoshort\ is
less than a third of the mass of the other two
clusters (Table~\ref{tab:optical_properties}).  Using the
\planck\ $\tilde{Y}_{500}$--$N_{200\overline{\rho}}$ relation, we
estimate $Y_{500} = 2.7\pm1.2 \times 10^{-6} \rm ~Mpc^2$, which
corresponds to $415 \pm 190~\mu$Jy of integrated SZE flux within
$R_{500}$ at 31~GHz.  This corresponds to an SZA signal smaller than
2$\sigma$ for the observations considered here, below the SZA detection
threshold.  Furthermore, the above significance calculation assumes
that the entire SZE flux within $R_{500}$ is actually contained within
the synthesized beam (effective resolution) of the SZA.  However, the
SZA synthesized beam is $1.\!\!\arcmin7$, smaller than $R_{500}$
($2.\!\!\arcmin6$) for this cluster, and the cluster SZE signal will
be diluted.

%%%%%
% Y500-N200 relation
%%%%%

%%%%%%%%%%%%%%%%%%%%%%%%%%%%%%%%%%%%%%%%%%%%%%%%%%%%%%%%%%%%%%%%%%%%%%%
%%%%%%%%%%%%%%%%%%%%%%%%%%%%%%%%%%%%%%%%%%%%%%%%%%%%%%%%%%%%%%%%%%%%%%%

Our mass estimates for \coneshort\ and \cthreeshort\ confirm that they
are massive systems ($M_{500} \simeq 10^{15}$ $M_\odot$).  This is
consistent with optical and X-ray follow-up of the initial cluster
results from ACT's southern region that suggests the high significance
detections will be $\gtrsim 8\times 10^{14}$ $M_\odot$
\citep{menanteau2010}.  The $\beta$ and A10 model fits to the SZA data
yield consistent masses for both clusters.  A comparison of the SZE,
X-ray, and joint mass estimates of \cthreeshort\ shows more scatter.
This scatter might be indicative of \cthreeshort\ being an unrelaxed
cluster, especially in light of the asymmetric structure seen in both
the SZA and \chandra\ observations (see Fig.~\ref{fig:a2631_x-sze}).
However, given the uncertainties in the masses, the different
methodologies yield consistent results for \cthreeshort\ as well.

\coneshort\ was discovered by ACT and we present initial mass
estimates but there are previous X-ray analyses of \cthreeshort.
Temperatures are computed from regions that differ between the
different works.  Using the earlier of the two \chandra\ observations,
\citet{maughan2008} finds $T_e = 6.5 \pm 0.6$~keV within
(0.15--1)$R_{500}$, where $R_{500}=1.2$~Mpc.  Analyses of \xmm\ data
using a number of methods for computing $R_{500}$ yields $T_e = 7.5
\pm 0.5$~keV (within (0.2--0.5)$R_{500}$) and an average $M_{500}=(8
\pm 2) \times 10^{14}\ M_\odot$ within $R_{500}=1.2$~Mpc
\citep{zhang2008}.  Another analysis of \xmm\ data finds $T_e = 7.7
\pm 0.6$~keV within an aperture containing 90\% of the
background-subtracted surface intensity \citep{andersson2009}.  Our
derived $T_e = 7.3 \pm 0.6$~keV agrees to within 1.3$\sigma$ with other
\chandra\ and \xmm\ measurements, despite the differences in radii.
The \citet{zhang2008} $M_{500}$ is consistent with our derived SZE and
X-ray masses to within 2$\sigma$ in the most discrepant case.  Despite
data from various observatories and different methodologies, the
temperatures and masses of \cthreeshort\ are consistent.

A number of systematics may affect these mass measurements, summarized
in Table~\ref{tab:systematics} as percentages for a single cluster.
Both simulation and analysis of X-ray data suggest that asphericity
typically affects mass estimates at the $5$--$10$\% level when
measured at large radii such as $R_{500}$
\citep[e.g.,][]{mathiesen1999, piffaretti2003}.  We conservatively
adopt $10$\% for the effects of asphericity.  Simulations suggest that
small-scale fluctuations, often called clumping, will cause a $\simeq
10$\% overestimate of mass \citep{mathiesen1999}.  Non-thermal
pressure support may affect both SZE and X-ray mass estimates, causing
mass to be underestimated at the 10-20\% level
\citep[e.g.,][]{nagai2007, lau2009}.  For the SZE-only analysis,
changes in the assumed gas fraction, on average, change the mass by
12\%, with lower gas fraction leading to higher mass; the fixed
metallicity assumption is a $\leq 1$\% effect \citep{mroczkowski2011},
which we neglect.  Relativistic corrections to the SZE are a 3\%
correction at 30~GHz for an 8~keV cluster like
\cthreeshort\ \citep{itoh1998}.

The SZA and \chandra\ calibrations both affect the mass estimates.
The SZE mass determinations depend on the SZE signal as $M\propto
(\Delta T)^{1/2}$ so that the conservative 20\% calibration results in
a 10\% change in mass.  Recent changes in the \chandra\ calibration
can change the cluster temperatures inferred from spectroscopic fits
by $\simeq 10$\% \citep{reese2010, nevalainen2010}.  This will directly
impact our X-ray mass estimates since $M \propto T_e$.  

Potential systematics are summarized in Table~\ref{tab:systematics}, with
totals added in quadrature.  Both SZE and X-ray estimates have
potential systematics at the $\simeq 20$\% level, roughly the same order
as the statistical uncertainties (Table~\ref{tab:mass}).

%%%%%
%
% Systematics
%
%%%%%

%% Cause   SZE   X-ray

\begin{deluxetable}{lcc}

%\singlespace
%\footnotesize
%\rotate
\tablewidth{0pt}
%\tablenum{}
\tablecolumns{3}
%\tableheadfrac{}
\tablecaption{Potential Systematics on Mass\label{tab:systematics}}
\tablehead{
\colhead{} &
\colhead{SZE} &
\colhead{X-ray}\\
\colhead{Source} &
\colhead{(\%)} &
\colhead{(\%)} 
}
\startdata
Calibration       & $\pm 10$ & $\pm 10$ \\
Asphericity       & $\pm 10$ & $\pm 10$ \\
Clumping          & $-10$    & $-10$    \\
Non-thermal $P_e$ & $+15$    & $+15$    \\
Assumed $f_g$     & $\pm 12$ & \nodata  \\
Relativisic corrections & $+3$& \nodata \\
\hline
\\[-0.5pc]
Total\tablenotemark{a} & $^{+24} _{-21}$  & $^{+21} _{-17}$
\enddata
\tablenotetext{a}{Added in quadrature.}
%\tablecomment{}
\end{deluxetable}

%%%%%%%%%%%%%%%%%%%%%%%%%%%%%%%%%%%%%%%%%%%%%%%%%%%%%%%%%%%%%%%%%%%%%%
\section{Conclusion} 
\label{sec:conclusion}
%%%%%%%%%%%%%%%%%%%%%%%%%%%%%%%%%%%%%%%%%%%%%%%%%%%%%%%%%%%%%%%%%%%%%%

We obtained SZA follow-up observations of three optically-confirmed
clusters from preliminary ACT maps along the celestial equator.
\cone\ is a massive, high-redshift cluster newly discovered by ACT.
The SZA detects two of the three clusters at high significance.  ACT
and SZA data show good qualitative agreement
(Figure~\ref{fig:low_res}).  The cluster \cthreeshort\ shows good
agreement between SZE and X-ray data (Figure~\ref{fig:a2631_x-sze})
with both peaks aligning and similar E-W elongation seen in both
wavebands.  Initial mass estimates confirm that \coneshort\ and
\cthreeshort\ are massive clusters with $M_{500} \simeq
10^{15}\ M_\odot$.

Two compact radio sources are detected by the SZA in each cluster
field (Table~\ref{tab:pt}).  Using NVSS 1.4~GHz flux densities, we
compute spectral indices and predict the flux densities in ACT's
148~GHz band.  The radio sources are expected to be $\lesssim 6$~mJy
in the ACT 148~GHz maps, suggesting that radio sources are not a
significant contaminant for detection of high mass clusters.  However,
they can still impact the measured SZE signal.  As an example, the
brighter source in \cthreeshort\ could be filling in the SZE signal at
the $\lesssim 20$\% level, assuming the extrapolation of the sources'
lower-frequency flux densities holds.  A more precise determination of
potential contamination of the SZE signal from compact sources at ACT
frequencies will be obtained through observing a larger number of ACT
cluster fields over a range of frequencies.

Optimized for different purposes, ACT and SZA provide complementary
SZE data on galaxy clusters.  With the ability to quickly image large
regions of the sky, ACT is well-suited to finding clusters.  Targeted
observations with the SZA allow deep integrations for detailed cluster
studies.  The SZA images (Fig.~\ref{fig:low_res}) show peak S/N of 12
and 10 for \coneshort\ and \cthreeshort, respectively, compared to
$\mbox{S/N} \simeq 6$ for the match-filtered ACT images.  The SZA
provides higher resolution than ACT, measuring angular scales
$15\arcsec$--$5\arcmin$, well-matched to clusters.  In addition, the
spatial filtering of the interferometer provides a clean method of
disentangling cluster emission from radio point source emission.

This pilot study of SZA follow-up observations of ACT-detected
clusters shows that the detected clusters are massive systems.
Cluster abundances are exponentially sensitive to mass
\citep[e.g.,][]{press1974}, with the most massive clusters providing
the most leverage on cosmological studies.  Finding extremely massive
systems at high redshift can potentially rule out the current
$\Lambda$-CDM paradigm \citep[e.g.,][]{mortonson2011}.  In addition,
cosmological determinations utilizing massive systems are expected to
be less susceptible to the effects of non-gravitational astrophysics.
The highest mass systems from ACT's two survey regions will comprise
our core sample for derivation of cosmological parameters from
clusters.  It is crucial to measure well the integrated SZE signal and
the masses of these clusters for proper cosmological interpretation of
cluster yields from current and future SZE cluster surveys.  Deep,
targeted SZA observations provide a method of disentangling the point
source and cluster emissions and enable robust estimates of the
integrated SZE signal, $Y_{500}$, and initial mass estimates.  When
combined with mass measures from X-ray and weak lensing, this will
provide a robust measure of the mass-SZE flux scaling relation.

\acknowledgements

We are grateful to John Carpenter for guidance on CARMA observing, the
queue, and a crash-course in Miriad.  We thank the CARMA
collaboration, especially Tom Plagge, for many discussions of the
nitty-gritty details of the data.  Support for TM was provided by NASA
through the Einstein Fellowship Program, grant PF0-110077.  The CARMA
3.5-m observations presented here were awarded in proposals c0563 and
c0619.

This work was supported by the U.S. National Science Foundation
through awards AST-0408698 for the ACT project, and PHY-0355328,
AST-0707731 and PIRE-0507768. Funding was also provided by Princeton
University and the University of Pennsylvania.  The PIRE program made
possible exchanges between Chile, South Africa, Spain and the US that
enabled this research program.  Computations were performed on the GPC
supercomputer at the SciNet HPC Consortium.  SciNet is funded by: the
Canada Foundation for Innovation under the auspices of Compute Canada;
the Government of Ontario; Ontario Research Fund -- Research
Excellence; and the University of Toronto.

Support for CARMA construction was derived from the Gordon and Betty
Moore Foundation, the Kenneth T. and Eileen L. Norris Foundation, the
James S. McDonnell Foundation, the Associates of the California
Institute of Technology, the University of Chicago, the states of
California, Illinois, and Maryland, and the National Science
Foundation. Ongoing CARMA development and operations are supported by
the National Science Foundation under a cooperative agreement, and by
the CARMA partner universities.

Funding for the SDSS and SDSS-II has been provided by the Alfred
P. Sloan Foundation, the Participating Institutions, the National
Science Foundation, the U.S. Department of Energy, the National
Aeronautics and Space Administration, the Japanese Monbukagakusho, the
Max Planck Society, and the Higher Education Funding Council for
England.  The SDSS is managed by the Astrophysical Research Consortium
for the Participating Institutions. The Participating Institutions are
the American Museum of Natural History, Astrophysical Institute
Potsdam, University of Basel, University of Cambridge, Case Western
Reserve University, University of Chicago, Drexel University,
Fermilab, the Institute for Advanced Study, the Japan Participation
Group, Johns Hopkins University, the Joint Institute for Nuclear
Astrophysics, the Kavli Institute for Particle Astrophysics and
Cosmology, the Korean Scientist Group, the Chinese Academy of Sciences
(LAMOST), Los Alamos National Laboratory, the Max-Planck-Institute for
Astronomy (MPIA), the Max-Planck-Institute for Astrophysics (MPA), New
Mexico State University, Ohio State University, University of
Pittsburgh, University of Portsmouth, Princeton University, the United
States Naval Observatory, and the University of Washington.

This work made use of observations obtained with the Apache Point
Observatory 3.5-meter telescope, which is owned and operated by the
Astrophysical Research Consortium and observations obtained at the
Gemini Observatory, which is operated by the Association of
Universities for Research in Astronomy, Inc., under a cooperative
agreement with the NSF on behalf of the Gemini partnership.

This research has made use of the NASA/IPAC Extragalactic Database
(NED) which is operated by the Jet Propulsion Laboratory, California
Institute of Technology, under contract with the National Aeronautics
and Space Administration.  This research has made use of data obtained
from the Chandra Data Archive and the CIAO software provided by the
Chandra X-ray Center (CXC).

%BiBTeX stuff
%\input{ms.bbl}
{
\bibliography{clusters}

\begin{thebibliography}{99}
\expandafter\ifx\csname natexlab\endcsname\relax\def\natexlab#1{#1}\fi

\bibitem[{{Abazajian} {et~al.}(2009){Abazajian}, {Adelman-McCarthy},
  {Ag{\"u}eros}, {Allam}, {Allende Prieto}, {An}, {Anderson}, {Anderson},
  {Annis}, {Bahcall}, \& et~al.}]{abazajian2009}
{Abazajian}, K.~N., et~al. 2009, \apjs, 182, 543

\bibitem[{{Andersson} {et~al.}(2009){Andersson}, {Peterson}, {Madejski}, \&
  {Goobar}}]{andersson2009}
{Andersson}, K., {Peterson}, J.~R., {Madejski}, G., \& {Goobar}, A. 2009, \apj,
  696, 1029

\bibitem[{{Arnaud}(1996)}]{arnaud1996}
{Arnaud}, K.~A. 1996, in Astronomical Society of the Pacific Conference Series,
  Vol. 101, Astronomical Data Analysis Software and Systems V, ed.
  {G.~H.~Jacoby \& J.~Barnes}, 17

\bibitem[{{Arnaud} {et~al.}(2010){Arnaud}, {Pratt}, {Piffaretti},
  {B{\"o}hringer}, {Croston}, \& {Pointecouteau}}]{arnaud2010}
{Arnaud}, M., {Pratt}, G.~W., {Piffaretti}, R., {B{\"o}hringer}, H., {Croston},
  J.~H., \& {Pointecouteau}, E. 2010, \aap, 517, A92

\bibitem[{{Asplund} {et~al.}(2009){Asplund}, {Grevesse}, {Sauval}, \&
  {Scott}}]{asplund2009}
{Asplund}, M., {Grevesse}, N., {Sauval}, A.~J., \& {Scott}, P. 2009, \araa, 47,
  481

\bibitem[{{Bahcall} {et~al.}(1997){Bahcall}, {Fan}, \& {Cen}}]{bahcall1997}
{Bahcall}, N.~A., {Fan}, X., \& {Cen}, R. 1997, \apjl, 485, L53

\bibitem[{{Balucinska-Church} \& {McCammon}(1992)}]{balucinska1992}
{Balucinska-Church}, M. \& {McCammon}, D. 1992, \apj, 400, 699

\bibitem[{{Bartlett} \& {Silk}(1994)}]{bartlett1994}
{Bartlett}, J.~G. \& {Silk}, J. 1994, \apj, 423, 12

\bibitem[{{Becker} {et~al.}(1995){Becker}, {White}, \& {Helfand}}]{becker1995}
{Becker}, R.~H., {White}, R.~L., \& {Helfand}, D.~J. 1995, \apj, 450, 559

\bibitem[{{Birkinshaw}(1999)}]{birkinshaw1999}
{Birkinshaw}, M. 1999, \physrep, 310, 97

\bibitem[{{Bonamente} {et~al.}(2004){Bonamente}, {Joy}, {Carlstrom}, {Reese},
  \& {LaRoque}}]{bonamente2004}
{Bonamente}, M., {Joy}, M.~K., {Carlstrom}, J.~E., {Reese}, E.~D., \&
  {LaRoque}, S.~J. 2004, \apj, 614, 56

\bibitem[{{Borgani} {et~al.}(2001){Borgani}, {Rosati}, {Tozzi}, {Stanford},
  {Eisenhardt}, {Lidman}, {Holden}, {Della Ceca}, {Norman}, \&
  {Squires}}]{borgani2001}
{Borgani}, S., et~al. 2001, \apj, 561, 13

\bibitem[{{Carlstrom} {et~al.}(2011){Carlstrom}, {Ade}, {Aird}, {Benson},
  {Bleem}, {Busetti}, {Chang}, {Chauvin}, {Cho}, {Crawford}, {Crites}, {Dobbs},
  {Halverson}, {Heimsath}, {Holzapfel}, {Hrubes}, {Joy}, {Keisler}, {Lanting},
  {Lee}, {Leitch}, {Leong}, {Lu}, {Lueker}, {Luong-Van}, {McMahon}, {Mehl},
  {Meyer}, {Mohr}, {Montroy}, {Padin}, {Plagge}, {Pryke}, {Ruhl}, {Schaffer},
  {Schwan}, {Shirokoff}, {Spieler}, {Staniszewski}, {Stark}, {Tucker},
  {Vanderlinde}, {Vieira}, \& {Williamson}}]{carlstrom2011}
{Carlstrom}, J.~E., et~al. 2011, \pasp, 123, 903

\bibitem[{{Carlstrom} {et~al.}(2002){Carlstrom}, {Holder}, \&
  {Reese}}]{carlstrom2002}
{Carlstrom}, J.~E., {Holder}, G.~P., \& {Reese}, E.~D. 2002, \araa, 40, 643

\bibitem[{{Cash}(1979)}]{cash1979}
{Cash}, W. 1979, \apj, 228, 939

\bibitem[{{Cavaliere} \& {Fusco-Femiano}(1976)}]{cavaliere1976}
{Cavaliere}, A. \& {Fusco-Femiano}, R. 1976, \aap, 49, 137

\bibitem[{{Cavaliere} \& {Fusco-Femiano}(1978)}]{cavaliere1978}
---. 1978, \aap, 70, 677

\bibitem[{{Challinor} \& {Lasenby}(1998)}]{challinor1998}
{Challinor}, A. \& {Lasenby}, A. 1998, \apj, 499, 1

\bibitem[{{Coble} {et~al.}(2007){Coble}, {Bonamente}, {Carlstrom}, {Dawson},
  {Hasler}, {Holzapfel}, {Joy}, {La Roque}, {Marrone}, \& {Reese}}]{coble2007}
{Coble}, K., {Bonamente}, M., {Carlstrom}, J.~E., {Dawson}, K., {Hasler}, N.,
  {Holzapfel}, W., {Joy}, M., {La Roque}, S., {Marrone}, D.~P., \& {Reese},
  E.~D. 2007, \aj, 134, 897

\bibitem[{{Condon} {et~al.}(1998){Condon}, {Cotton}, {Greisen}, {Yin},
  {Perley}, {Taylor}, \& {Broderick}}]{condon1998}
{Condon}, J.~J., {Cotton}, W.~D., {Greisen}, E.~W., {Yin}, Q.~F., {Perley},
  R.~A., {Taylor}, G.~B., \& {Broderick}, J.~J. 1998, \aj, 115, 1693

\bibitem[{{Cooray} {et~al.}(1998){Cooray}, {Grego}, {Holzapfel}, {Joy}, \&
  {Carlstrom}}]{cooray1998a}
{Cooray}, A.~R., {Grego}, L., {Holzapfel}, W.~L., {Joy}, M., \& {Carlstrom},
  J.~E. 1998, \aj, 115, 1388

\bibitem[{{Das} {et~al.}(2011){Das}, {Marriage}, {Ade}, {Aguirre}, {Amiri},
  {Appel}, {Barrientos}, {Battistelli}, {Bond}, {Brown}, {Burger}, {Chervenak},
  {Devlin}, {Dicker}, {Bertrand Doriese}, {Dunkley}, {D{\"u}nner},
  {Essinger-Hileman}, {Fisher}, {Fowler}, {Hajian}, {Halpern}, {Hasselfield},
  {Hern{\'a}ndez-Monteagudo}, {Hilton}, {Hilton}, {Hincks}, {Hlozek},
  {Huffenberger}, {Hughes}, {Hughes}, {Infante}, {Irwin}, {Baptiste Juin},
  {Kaul}, {Klein}, {Kosowsky}, {Lau}, {Limon}, {Lin}, {Lupton}, {Marsden},
  {Martocci}, {Mauskopf}, {Menanteau}, {Moodley}, {Moseley}, {Netterfield},
  {Niemack}, {Nolta}, {Page}, {Parker}, {Partridge}, {Reid}, {Sehgal},
  {Sherwin}, {Sievers}, {Spergel}, {Staggs}, {Swetz}, {Switzer}, {Thornton},
  {Trac}, {Tucker}, {Warne}, {Wollack}, \& {Zhao}}]{das2011}
{Das}, S., et~al.  2011, \apj, 729, 62

\bibitem[{{Dorman} \& {Arnaud}(2001)}]{dorman2001}
{Dorman}, B. \& {Arnaud}, K.~A. 2001, in Astronomical Society of the Pacific
  Conference Series, Vol. 238, Astronomical Data Analysis Software and Systems
  X, ed. {F.~R.~Harnden Jr., F.~A.~Primini, \& H.~E.~Payne}, 415

\bibitem[{{Dunkley} {et~al.}(2010){Dunkley}, {Hlozek}, {Sievers}, {Acquaviva},
  {Ade}, {Aguirre}, {Amiri}, {Appel}, {Barrientos}, {Battistelli}, {Bond},
  {Brown}, {Burger}, {Chervenak}, {Das}, {Devlin}, {Dicker}, {Bertrand
  Doriese}, {Dunner}, {Essinger-Hileman}, {Fisher}, {Fowler}, {Hajian},
  {Halpern}, {Hasselfield}, {Hernandez-Monteagudo}, {Hilton}, {Hilton},
  {Hincks}, {Huffenberger}, {Hughes}, {Hughes}, {Infante}, {Irwin}, {Juin},
  {Kaul}, {Klein}, {Kosowsky}, {Lau}, {Limon}, {Lin}, {Lupton}, {Marriage},
  {Marsden}, {Mauskopf}, {Menanteau}, {Moodley}, {Moseley}, {Netterfield},
  {Niemack}, {Nolta}, {Page}, {Parker}, {Partridge}, {Reid}, {Sehgal},
  {Sherwin}, {Spergel}, {Staggs}, {Swetz}, {Switzer}, {Thornton}, {Trac},
  {Tucker}, {Warne}, {Wollack}, \& {Zhao}}]{dunkley2011}
  {Dunkley}, J., et~al.  2010, arXiv:1009.0866

\bibitem[{{Eke} {et~al.}(1998){Eke}, {Cole}, {Frenk}, \& {Patrick
  Henry}}]{eke1998}
{Eke}, V.~R., {Cole}, S., {Frenk}, C.~S., \& {Patrick Henry}, J. 1998, \mnras,
  298, 1145

\bibitem[{{Fowler} {et~al.}(2010){Fowler}, {Acquaviva}, {Ade}, {Aguirre},
  {Amiri}, {Appel}, {Barrientos}, {Battistelli}, {Bond}, {Brown}, {Burger},
  {Chervenak}, {Das}, {Devlin}, {Dicker}, {Doriese}, {Dunkley}, {D{\"u}nner},
  {Essinger-Hileman}, {Fisher}, {Hajian}, {Halpern}, {Hasselfield},
  {Hern{\'a}ndez-Monteagudo}, {Hilton}, {Hilton}, {Hincks}, {Hlozek},
  {Huffenberger}, {Hughes}, {Hughes}, {Infante}, {Irwin}, {Jimenez}, {Juin},
  {Kaul}, {Klein}, {Kosowsky}, {Lau}, {Limon}, {Lin}, {Lupton}, {Marriage},
  {Marsden}, {Martocci}, {Mauskopf}, {Menanteau}, {Moodley}, {Moseley},
  {Netterfield}, {Niemack}, {Nolta}, {Page}, {Parker}, {Partridge}, {Quintana},
  {Reid}, {Sehgal}, {Sievers}, {Spergel}, {Staggs}, {Swetz}, {Switzer},
  {Thornton}, {Trac}, {Tucker}, {Verde}, {Warne}, {Wilson}, {Wollack}, \&
  {Zhao}}]{fowler2010}
{Fowler}, J.~W., et~al.  2010, \apj, 722, 1148

\bibitem[{{Fowler} {et~al.}(2007){Fowler}, {Niemack}, {Dicker}, {Aboobaker},
  {Ade}, {Battistelli}, {Devlin}, {Fisher}, {Halpern}, {Hargrave}, {Hincks},
  {Kaul}, {Klein}, {Lau}, {Limon}, {Marriage}, {Mauskopf}, {Page}, {Staggs},
  {Swetz}, {Switzer}, {Thornton}, \& {Tucker}}]{fowler2007}
{Fowler}, J.~W., et~al.  2007, \ao, 46, 3444

\bibitem[{{Geweke}(1992)}]{geweke1992}
{Geweke}, J. 1992, in Bayesian Statitstics IV, ed. J.~M.~B. et~al.
  (Oxford:~Clarendon), 169

\bibitem[{{Haehnelt} \& {Tegmark}(1996)}]{haehnelt1996}
{Haehnelt}, M.~G. \& {Tegmark}, M. 1996, \mnras, 279, 545

\bibitem[{{Haiman} {et~al.}(2001){Haiman}, {Mohr}, \& {Holder}}]{haiman2001}
{Haiman}, Z., {Mohr}, J.~J., \& {Holder}, G.~P. 2001, \apj, 553, 545

\bibitem[{{Hajian} {et~al.}(2010){Hajian}, {Acquaviva}, {Ade}, {Aguirre},
  {Amiri}, {Appel}, {Barrientos}, {Battistelli}, {Bond}, {Brown}, {Burger},
  {Chervenak}, {Das}, {Devlin}, {Dicker}, {Bertrand Doriese}, {Dunkley},
  {Dunner}, {Essinger-Hileman}, {Fisher}, {Fowler}, {Halpern}, {Hasselfield},
  {Hernandez-Monteagudo}, {Hilton}, {Hilton}, {Hincks}, {Hlozek},
  {Huffenberger}, {Hughes}, {Hughes}, {Infante}, {Irwin}, {Baptiste Juin},
  {Kaul}, {Klein}, {Kosowsky}, {Lau}, {Limon}, {Lin}, {Lupton}, {Marriage},
  {Marsden}, {Mauskopf}, {Menanteau}, {Moodley}, {Moseley}, {Netterfield},
  {Niemack}, {Nolta}, {Page}, {Parker}, {Partridge}, {Reid}, {Sehgal},
  {Sherwin}, {Sievers}, {Spergel}, {Staggs}, {Swetz}, {Switzer}, {Thornton},
  {Trac}, {Tucker}, {Warne}, {Wollack}, \& {Zhao}}]{hajian2011}
{Hajian}, A., et~al.  2010, arXiv:1009.0777

\bibitem[{{Henry}(2004)}]{henry2004}
{Henry}, J.~P. 2004, \apj, 609, 603

\bibitem[{{Henry} \& {Arnaud}(1991)}]{henry1991}
{Henry}, J.~P. \& {Arnaud}, K.~A. 1991, \apj, 372, 410

\bibitem[{{Herranz} {et~al.}(2002){Herranz}, {Sanz}, {Barreiro}, {Hobson},
  {Martinez-Gonzalez}, \& {Diego}}]{herranz2002}
{Herranz}, D., {Sanz}, J.~L., {Barreiro}, R.~B., {Hobson}, M.,
  {Martinez-Gonzalez}, E., \& {Diego}, J.~M. 2002, in Society of Photo-Optical
  Instrumentation Engineers (SPIE) Conference Series, Vol. 4847, Society of
  Photo-Optical Instrumentation Engineers (SPIE) Conference Series, ed.
  {J.-L.~Starck \& F.~D.~Murtagh}, 50

\bibitem[{{Holder} {et~al.}(2000){Holder}, {Mohr}, {Carlstrom}, {Evrard}, \&
  {Leitch}}]{holder2000}
{Holder}, G.~P., {Mohr}, J.~J., {Carlstrom}, J.~E., {Evrard}, A.~E., \&
  {Leitch}, E.~M. 2000, \apj, 544, 629

\bibitem[{{Itoh} {et~al.}(1998){Itoh}, {Kohyama}, \& {Nozawa}}]{itoh1998}
{Itoh}, N., {Kohyama}, Y., \& {Nozawa}, S. 1998, \apj, 502, 7

\bibitem[{{Johnston} {et~al.}(2007){Johnston}, {Sheldon}, {Tasitsiomi},
  {Frieman}, {Wechsler}, \& {McKay}}]{johnston2007}
{Johnston}, D.~E., {Sheldon}, E.~S., {Tasitsiomi}, A., {Frieman}, J.~A.,
  {Wechsler}, R.~H., \& {McKay}, T.~A. 2007, \apj, 656, 27

\bibitem[{{Kalberla} {et~al.}(2005){Kalberla}, {Burton}, {Hartmann}, {Arnal},
  {Bajaja}, {Morras}, \& {P{\"o}ppel}}]{kalberla2005}
{Kalberla}, P.~M.~W., {Burton}, W.~B., {Hartmann}, D., {Arnal}, E.~M.,
  {Bajaja}, E., {Morras}, R., \& {P{\"o}ppel}, W.~G.~L. 2005, \aap, 440, 775

\bibitem[{{Koester} {et~al.}(2007){Koester}, {McKay}, {Annis}, {Wechsler},
  {Evrard}, {Bleem}, {Becker}, {Johnston}, {Sheldon}, {Nichol}, {Miller},
  {Scranton}, {Bahcall}, {Barentine}, {Brewington}, {Brinkmann}, {Harvanek},
  {Kleinman}, {Krzesinski}, {Long}, {Nitta}, {Schneider}, {Sneddin}, {Voges},
  \& {York}}]{koester2007}
{Koester}, B.~P., et~al.  2007, \apj, 660, 239

\bibitem[{{Komatsu} {et~al.}(2009){Komatsu}, {Dunkley}, {Nolta}, {Bennett},
  {Gold}, {Hinshaw}, {Jarosik}, {Larson}, {Limon}, {Page}, {Spergel},
  {Halpern}, {Hill}, {Kogut}, {Meyer}, {Tucker}, {Weiland}, {Wollack}, \&
  {Wright}}]{komatsu2009}
{Komatsu}, E., et~al.  2009,  \apjs, 180, 330

\bibitem[{{Komatsu} {et~al.}(2011){Komatsu}, {Smith}, {Dunkley}, {Bennett},
  {Gold}, {Hinshaw}, {Jarosik}, {Larson}, {Nolta}, {Page}, {Spergel},
  {Halpern}, {Hill}, {Kogut}, {Limon}, {Meyer}, {Odegard}, {Tucker}, {Weiland},
  {Wollack}, \& {Wright}}]{komatsu2011}
{Komatsu}, E., et~al.  2011, \apjs, 192, 18

\bibitem[{{Lau} {et~al.}(2009){Lau}, {Kravtsov}, \& {Nagai}}]{lau2009}
{Lau}, E.~T., {Kravtsov}, A.~V., \& {Nagai}, D. 2009, \apj, 705, 1129

\bibitem[{{Lin} \& {Mohr}(2007)}]{lin2007}
{Lin}, Y. \& {Mohr}, J.~J. 2007, \apjs, 170, 71

\bibitem[{{Lin} {et~al.}(2009){Lin}, {Partridge}, {Pober}, {Bouchefry},
  {Burke}, {Klein}, {Coish}, \& {Huffenberger}}]{lin2009}
{Lin}, Y., {Partridge}, B., {Pober}, J.~C., {Bouchefry}, K.~E., {Burke}, S.,
  {Klein}, J.~N., {Coish}, J.~W., \& {Huffenberger}, K.~M. 2009, \apj, 694, 992

\bibitem[{{Majumdar} \& {Mohr}(2004)}]{majumdar2004}
{Majumdar}, S. \& {Mohr}, J.~J. 2004, \apj, 613, 41

\bibitem[{{Mantz} {et~al.}(2008){Mantz}, {Allen}, {Ebeling}, \&
  {Rapetti}}]{mantz2008}
{Mantz}, A., {Allen}, S.~W., {Ebeling}, H., \& {Rapetti}, D. 2008, \mnras, 387,
  1179

\bibitem[{{Mantz} {et~al.}(2010){Mantz}, {Allen}, {Rapetti}, \&
  {Ebeling}}]{mantz2010}
{Mantz}, A., {Allen}, S.~W., {Rapetti}, D., \& {Ebeling}, H. 2010, \mnras, 406,
  1759

\bibitem[{{Marriage} {et~al.}(2010){Marriage}, {Acquaviva}, {Ade}, {Aguirre},
  {Amiri}, {Appel}, {Barrientos}, {Battistelli}, {Bond}, {Brown}, {Burger},
  {Chervenak}, {Das}, {Devlin}, {Dicker}, {Doriese}, {Dunkley}, {Dunner},
  {Essinger-Hileman}, {Fisher}, {Fowler}, {Hajian}, {Halpern}, {Hasselfield},
  {Hern'andez-Monteagudo}, {Hilton}, {Hilton}, {Hincks}, {Hlozek},
  {Huffenberger}, {Hughes}, {Hughes}, {Infante}, {Irwin}, {Juin}, {Kaul},
  {Klein}, {Kosowsky}, {Lau}, {Limon}, {Lin}, {Lupton}, {Marsden}, {Martocci},
  {Mauskopf}, {Menanteau}, {Moodley}, {Moseley}, {Netterfield}, {Niemack},
  {Nolta}, {Page}, {Parker}, {Partridge}, {Quintana}, {Reese}, {Reid},
  {Sehgal}, {Sherwin}, {Sievers}, {Spergel}, {Staggs}, {Swetz}, {Switzer},
  {Thornton}, {Trac}, {Tucker}, {Warne}, {Wilson}, {Wollack}, \&
  {Zhao}}]{marriage2011b}
  {Marriage}, T.~A., et~al.  2010, arXiv:1010.1065

\bibitem[{{Marriage} {et~al.}(2011){Marriage}, {Baptiste Juin}, {Lin},
  {Marsden}, {Nolta}, {Partridge}, {Ade}, {Aguirre}, {Amiri}, {Appel},
  {Barrientos}, {Battistelli}, {Bond}, {Brown}, {Burger}, {Chervenak}, {Das},
  {Devlin}, {Dicker}, {Bertrand Doriese}, {Dunkley}, {D{\"u}nner},
  {Essinger-Hileman}, {Fisher}, {Fowler}, {Hajian}, {Halpern}, {Hasselfield},
  {Hern{\'a}ndez-Monteagudo}, {Hilton}, {Hilton}, {Hincks}, {Hlozek},
  {Huffenberger}, {Handel Hughes}, {Hughes}, {Infante}, {Irwin}, {Kaul},
  {Klein}, {Kosowsky}, {Lau}, {Limon}, {Lupton}, {Martocci}, {Mauskopf},
  {Menanteau}, {Moodley}, {Moseley}, {Netterfield}, {Niemack}, {Page},
  {Parker}, {Quintana}, {Reid}, {Sehgal}, {Sherwin}, {Sievers}, {Spergel},
  {Staggs}, {Swetz}, {Switzer}, {Thornton}, {Trac}, {Tucker}, {Warne},
  {Wilson}, {Wollack}, \& {Zhao}}]{marriage2011}
{Marriage}, T.~A., et~al. 2011, \apj, 731, 100

\bibitem[{{Massardi} \& {De Zotti}(2004)}]{massardi2004}
{Massardi}, M. \& {De Zotti}, G. 2004, \aap, 424, 409

\bibitem[{{Mathiesen} {et~al.}(1999){Mathiesen}, {Evrard}, \&
  {Mohr}}]{mathiesen1999}
{Mathiesen}, B., {Evrard}, A.~E., \& {Mohr}, J.~J. 1999, \apjl, 520, L21

\bibitem[{{Maughan} {et~al.}(2008){Maughan}, {Jones}, {Forman}, \& {Van
  Speybroeck}}]{maughan2008}
{Maughan}, B.~J., {Jones}, C., {Forman}, W., \& {Van Speybroeck}, L. 2008,
  \apjs, 174, 117

\bibitem[{{Mazzotta} {et~al.}(2004){Mazzotta}, {Rasia}, {Moscardini}, \&
  {Tormen}}]{mazzotta2004}
{Mazzotta}, P., {Rasia}, E., {Moscardini}, L., \& {Tormen}, G. 2004, \mnras,
  354, 10

\bibitem[{{Melin} {et~al.}(2006){Melin}, {Bartlett}, \&
  {Delabrouille}}]{melin2006}
{Melin}, J., {Bartlett}, J.~G., \& {Delabrouille}, J. 2006, \aap, 459, 341

\bibitem[{{Menanteau} {et~al.}(2010){Menanteau}, {Gonz{\'a}lez}, {Juin},
  {Marriage}, {Reese}, {Acquaviva}, {Aguirre}, {Appel}, {Baker}, {Barrientos},
  {Battistelli}, {Bond}, {Das}, {Deshpande}, {Devlin}, {Dicker}, {Dunkley},
  {D{\"u}nner}, {Essinger-Hileman}, {Fowler}, {Hajian}, {Halpern},
  {Hasselfield}, {Hern{\'a}ndez-Monteagudo}, {Hilton}, {Hincks}, {Hlozek},
  {Huffenberger}, {Hughes}, {Infante}, {Irwin}, {Klein}, {Kosowsky}, {Lin},
  {Marsden}, {Moodley}, {Niemack}, {Nolta}, {Page}, {Parker}, {Partridge},
  {Sehgal}, {Sievers}, {Spergel}, {Staggs}, {Swetz}, {Switzer}, {Thornton},
  {Trac}, {Warne}, \& {Wollack}}]{menanteau2010}
{Menanteau}, F., et~al.  2010, \apj, 723, 1523

\bibitem[{{Menanteau} {et~al.}(2010{\natexlab{b}}){Menanteau}, {Hughes},
  {Barrientos}, {Deshpande}, {Hilton}, {Infante}, {Jimenez}, {Kosowsky},
  {Moodley}, {Spergel}, \& {Verde}}]{menanteau2010b}
{Menanteau}, F., et~al. 2010{\natexlab{b}}, \apjs, 191, 340

\bibitem[{{Mortonson} {et~al.}(2011){Mortonson}, {Hu}, \&
  {Huterer}}]{mortonson2011}
{Mortonson}, M.~J., {Hu}, W., \& {Huterer}, D. 2011, \prd, 83, 023015

\bibitem[{{Mroczkowski}(2011)}]{mroczkowski2011}
{Mroczkowski}, T. 2011, \apjl, 728, L35

\bibitem[{{Mroczkowski} {et~al.}(2009){Mroczkowski}, {Bonamente}, {Carlstrom},
  {Culverhouse}, {Greer}, {Hawkins}, {Hennessy}, {Joy}, {Lamb}, {Leitch},
  {Loh}, {Maughan}, {Marrone}, {Miller}, {Muchovej}, {Nagai}, {Pryke}, {Sharp},
  \& {Woody}}]{mroczkowski2009}
{Mroczkowski}, T., et~al.  2009, \apj, 694, 1034

\bibitem[{{Muchovej} {et~al.}(2007){Muchovej}, {Mroczkowski}, {Carlstrom},
  {Cartwright}, {Greer}, {Hennessy}, {Loh}, {Pryke}, {Reddall}, {Runyan},
  {Sharp}, {Hawkins}, {Lamb}, {Woody}, {Joy}, {Leitch}, \&
  {Miller}}]{muchovej2007}
{Muchovej}, S., et~al.  2007, \apj, 663, 708

\bibitem[{{Nagai} {et~al.}(2007){Nagai}, {Kravtsov}, \&
  {Vikhlinin}}]{nagai2007}
{Nagai}, D., {Kravtsov}, A.~V., \& {Vikhlinin}, A. 2007, \apj, 668, 1

\bibitem[{{Navarro} {et~al.}(1996){Navarro}, {Frenk}, \& {White}}]{navarro1996}
{Navarro}, J.~F., {Frenk}, C.~S., \& {White}, S. D.~M. 1996, \apj, 462, 563

\bibitem[{{Navarro} {et~al.}(1997){Navarro}, {Frenk}, \& {White}}]{navarro1997}
---. 1997, \apj, 490, 493

\bibitem[{{Nevalainen} {et~al.}(2010){Nevalainen}, {David}, \&
  {Guainazzi}}]{nevalainen2010}
{Nevalainen}, J., {David}, L., \& {Guainazzi}, M. 2010, \aap, 523, A22

\bibitem[{{Piffaretti} {et~al.}(2003){Piffaretti}, {Jetzer}, \&
  {Schindler}}]{piffaretti2003}
{Piffaretti}, R., {Jetzer}, P., \& {Schindler}, S. 2003, \aap, 398, 41

\bibitem[{{Plagge} {et~al.}(2010){Plagge}, {Benson}, {Ade}, {Aird}, {Bleem},
  {Carlstrom}, {Chang}, {Cho}, {Crawford}, {Crites}, {de Haan}, {Dobbs},
  {George}, {Hall}, {Halverson}, {Holder}, {Holzapfel}, {Hrubes}, {Joy},
  {Keisler}, {Knox}, {Lee}, {Leitch}, {Lueker}, {Marrone}, {McMahon}, {Mehl},
  {Meyer}, {Mohr}, {Montroy}, {Padin}, {Pryke}, {Reichardt}, {Ruhl},
  {Schaffer}, {Shaw}, {Shirokoff}, {Spieler}, {Stalder}, {Staniszewski},
  {Stark}, {Vanderlinde}, {Vieira}, {Williamson}, \& {Zahn}}]{plagge2010}
{Plagge}, T., et~al. 2010,  \apj, 716, 1118

\bibitem[{{Planck Collaboration} (2011{\natexlab{a}}){Planck
  Collaboration}, {Ade}, {Aghanim}, {Arnaud}, {Ashdown}, {Aumont},
  {Baccigalupi}, {Baker}, {Balbi}, {Banday}, \& et~al.}]{planck2011a}
{Planck Collaboration}, {Ade}, P.~A.~R., et~al.  2011{\natexlab{a}},
arXiv:1101.2022

\bibitem[{{Planck Collaboration} (2011{\natexlab{b}}){Planck
  Collaboration}, {Ade}, {Aghanim}, {Arnaud}, {Ashdown}, {Aumont},
  {Baccigalupi}, {Balbi}, {Banday}, {Barreiro}, \& et~al.}]{planck2011d}
{Planck Collaboration}, {Ade}, P.~A.~R., et~al. 2011{\natexlab{b}}, arXiv:1101.2024

\bibitem[{{Planck Collaboration} (2011{\natexlab{c}}){Planck
  Collaboration}, {Aghanim}, {Arnaud}, {Ashdown}, {Aumont}, {Baccigalupi},
  {Balbi}, {Banday}, {Barreiro}, {Bartelmann}, \& et~al.}]{planck2011h}
{Planck Collaboration}, {Aghanim}, N., et~al. 2011{\natexlab{c}}, arXiv:1101.2027

\bibitem[{{Press} \& {Schechter}(1974)}]{press1974}
{Press}, W.~H. \& {Schechter}, P. 1974, \apj, 187, 425

\bibitem[{{Reese} {et~al.}(2002){Reese}, {Carlstrom}, {Joy}, {Mohr}, {Grego},
  \& {Holzapfel}}]{reese2002}
{Reese}, E.~D., {Carlstrom}, J.~E., {Joy}, M., {Mohr}, J.~J., {Grego}, L., \&
  {Holzapfel}, W.~L. 2002, \apj, 581, 53

\bibitem[{{Reese} {et~al.}(2010){Reese}, {Kawahara}, {Kitayama}, {Ota},
  {Sasaki}, \& {Suto}}]{reese2010}
{Reese}, E.~D., {Kawahara}, H., {Kitayama}, T., {Ota}, N., {Sasaki}, S., \&
  {Suto}, Y. 2010, \apj, 721, 653

\bibitem[{{Reese} {et~al.}(2000){Reese}, {Mohr}, {Carlstrom}, {Joy}, {Grego},
  {Holder}, {Holzapfel}, {Hughes}, {Patel}, \& {Donahue}}]{reese2000}
{Reese}, E.~D., {Mohr}, J.~J., {Carlstrom}, J.~E., {Joy}, M., {Grego}, L.,
  {Holder}, G.~P., {Holzapfel}, W.~L., {Hughes}, J.~P., {Patel}, S.~K., \&
  {Donahue}, M. 2000, \apj, 533, 38

\bibitem[{{Reiprich} \& {B{\"o}hringer}(2002)}]{reiprich2002}
{Reiprich}, T.~H. \& {B{\"o}hringer}, H. 2002, \apj, 567, 716

\bibitem[{{Rephaeli}(1995)}]{rephaeli1995}
{Rephaeli}, Y. 1995, \apj, 445, 33

\bibitem[{{Reyes} {et~al.}(2008){Reyes}, {Mandelbaum}, {Hirata}, {Bahcall}, \&
  {Seljak}}]{reyes2008}
{Reyes}, R., {Mandelbaum}, R., {Hirata}, C., {Bahcall}, N., \& {Seljak}, U.
  2008, \mnras, 390, 1157

\bibitem[{{Rozo} {et~al.}(2009){Rozo}, {Rykoff}, {Evrard}, {Becker}, {McKay},
  {Wechsler}, {Koester}, {Hao}, {Hansen}, {Sheldon}, {Johnston}, {Annis}, \&
  {Frieman}}]{rozo2009}
{Rozo}, E., et~al.  2009, \apj, 699, 768

\bibitem[{{Rozo} {et~al.}(2010){Rozo}, {Wechsler}, {Rykoff}, {Annis}, {Becker},
  {Evrard}, {Frieman}, {Hansen}, {Hao}, {Johnston}, {Koester}, {McKay},
  {Sheldon}, \& {Weinberg}}]{rozo2010}
{Rozo}, E., et~al.  2010, \apj, 708, 645

\bibitem[{{Sarazin}(1988)}]{sarazin1988}
{Sarazin}, C.~L. 1988, X-ray emission from clusters of galaxies (Cambridge
  Astrophysics Series, Cambridge: Cambridge University Press, 1988)

\bibitem[{{Sault} {et~al.}(1995){Sault}, {Teuben}, \& {Wright}}]{sault1995}
{Sault}, R.~J., {Teuben}, P.~J., \& {Wright}, M. C.~H. 1995, ASP Conf. Ser. 77:
  Astronomical Data Analysis Software and Systems IV, 4, 433

\bibitem[{{Schuecker} {et~al.}(2003){Schuecker}, {B{\"o}hringer}, {Collins}, \&
  {Guzzo}}]{schuecker2003}
{Schuecker}, P., {B{\"o}hringer}, H., {Collins}, C.~A., \& {Guzzo}, L. 2003,
  \aap, 398, 867

\bibitem[{{Sehgal} {et~al.}(2010){Sehgal}, {Bode}, {Das},
  {Hernandez-Monteagudo}, {Huffenberger}, {Lin}, {Ostriker}, \&
  {Trac}}]{sehgal2010}
{Sehgal}, N., {Bode}, P., {Das}, S., {Hernandez-Monteagudo}, C.,
  {Huffenberger}, K., {Lin}, Y.-T., {Ostriker}, J.~P., \& {Trac}, H. 2010,
  \apj, 709, 920

\bibitem[{{Sehgal} {et~al.}(2011){Sehgal}, {Trac}, {Acquaviva}, {Ade},
  {Aguirre}, {Amiri}, {Appel}, {Barrientos}, {Battistelli}, {Bond}, {Brown},
  {Burger}, {Chervenak}, {Das}, {Devlin}, {Dicker}, {Bertrand Doriese},
  {Dunkley}, {D{\"u}nner}, {Essinger-Hileman}, {Fisher}, {Fowler}, {Hajian},
  {Halpern}, {Hasselfield}, {Hern{\'a}ndez-Monteagudo}, {Hilton}, {Hilton},
  {Hincks}, {Hlozek}, {Holtz}, {Huffenberger}, {Hughes}, {Hughes}, {Infante},
  {Irwin}, {Jones}, {Baptiste Juin}, {Klein}, {Kosowsky}, {Lau}, {Limon},
  {Lin}, {Lupton}, {Marriage}, {Marsden}, {Martocci}, {Mauskopf}, {Menanteau},
  {Moodley}, {Moseley}, {Netterfield}, {Niemack}, {Nolta}, {Page}, {Parker},
  {Partridge}, {Reid}, {Sherwin}, {Sievers}, {Spergel}, {Staggs}, {Swetz},
  {Switzer}, {Thornton}, {Tucker}, {Warne}, {Wollack}, \& {Zhao}}]{sehgal2011}
{Sehgal}, N., et~al.  2011, \apj, 732, 44

\bibitem[{{Shepherd}(1997)}]{shepherd1997}
{Shepherd}, M.~C. 1997, in Astronomical Society of the Pacific Conference
  Series, Vol. 125, Astronomical Data Analysis Software and Systems VI, ed.
  {G.~Hunt \& H.~Payne}, 77

\bibitem[{{Smith} {et~al.}(2001){Smith}, {Brickhouse}, {Liedahl}, \&
  {Raymond}}]{smith2001}
{Smith}, R.~K., {Brickhouse}, N.~S., {Liedahl}, D.~A., \& {Raymond}, J.~C.
  2001, \apjl, 556, L91

\bibitem[{{Struble} \& {Rood}(1999)}]{struble1999}
{Struble}, M.~F. \& {Rood}, H.~J. 1999, \apjs, 125, 35

\bibitem[{{Sunyaev} \& {Zeldovich}(1980)}]{sunyaev1980}
{Sunyaev}, R.~A. \& {Zeldovich}, I.~B. 1980, \araa, 18, 537

\bibitem[{{Sunyaev} \& {Zel'dovich}(1970)}]{sunyaev1970}
{Sunyaev}, R.~A. \& {Zel'dovich}, Y.~B. 1970, Comments Astrophys. Space Phys.,
  2, 66

\bibitem[{{Sunyaev} \& {Zel'dovich}(1972)}]{sunyaev1972}
---. 1972, Comments Astrophys. Space Phys., 4, 173

\bibitem[{{Swetz} {et~al.}(2011){Swetz}, {Ade}, {Amiri}, {Appel},
  {Battistelli}, {Burger}, {Chervenak}, {Devlin}, {Dicker}, {Doriese},
  {D{\"u}nner}, {Essinger-Hileman}, {Fisher}, {Fowler}, {Halpern},
  {Hasselfield}, {Hilton}, {Hincks}, {Irwin}, {Jarosik}, {Kaul}, {Klein},
  {Lau}, {Limon}, {Marriage}, {Marsden}, {Martocci}, {Mauskopf}, {Moseley},
  {Netterfield}, {Niemack}, {Nolta}, {Page}, {Parker}, {Staggs}, {Stryzak},
  {Switzer}, {Thornton}, {Tucker}, {Wollack}, \& {Zhao}}]{swetz2011}
{Swetz}, D.~S., et~al.  2011,  \apjs, 194, 41

\bibitem[{{Vanderlinde} {et~al.}(2010){Vanderlinde}, {Crawford}, {de Haan},
  {Dudley}, {Shaw}, {Ade}, {Aird}, {Benson}, {Bleem}, {Brodwin}, {Carlstrom},
  {Chang}, {Crites}, {Desai}, {Dobbs}, {Foley}, {George}, {Gladders}, {Hall},
  {Halverson}, {High}, {Holder}, {Holzapfel}, {Hrubes}, {Joy}, {Keisler},
  {Knox}, {Lee}, {Leitch}, {Loehr}, {Lueker}, {Marrone}, {McMahon}, {Mehl},
  {Meyer}, {Mohr}, {Montroy}, {Ngeow}, {Padin}, {Plagge}, {Pryke}, {Reichardt},
  {Rest}, {Ruel}, {Ruhl}, {Schaffer}, {Shirokoff}, {Song}, {Spieler},
  {Stalder}, {Staniszewski}, {Stark}, {Stubbs}, {van Engelen}, {Vieira},
  {Williamson}, {Yang}, {Zahn}, \& {Zenteno}}]{vanderlinde2010}
{Vanderlinde}, K., et~al.  2010, \apj, 722, 1180

\bibitem[{{Viana} \& {Liddle}(1996)}]{viana1996}
{Viana}, P.~T.~P. \& {Liddle}, A.~R. 1996, \mnras, 281, 323

\bibitem[{{Vikhlinin} {et~al.}(2006){Vikhlinin}, {Kravtsov}, {Forman}, {Jones},
  {Markevitch}, {Murray}, \& {Van Speybroeck}}]{vikhlinin2006b}
{Vikhlinin}, A., {Kravtsov}, A., {Forman}, W., {Jones}, C., {Markevitch}, M.,
  {Murray}, S.~S., \& {Van Speybroeck}, L. 2006, \apj, 640, 691

\bibitem[{{Vikhlinin} {et~al.}(2009){Vikhlinin}, {Kravtsov}, {Burenin},
  {Ebeling}, {Forman}, {Hornstrup}, {Jones}, {Murray}, {Nagai}, {Quintana}, \&
  {Voevodkin}}]{vikhlinin2009b}
{Vikhlinin}, A., et~al. 2009, \apj, 692, 1060

\bibitem[{{Vikhlinin} {et~al.}(2005){Vikhlinin}, {Markevitch}, {Murray},
  {Jones}, {Forman}, \& {Van Speybroeck}}]{vikhlinin2005}
{Vikhlinin}, A., {Markevitch}, M., {Murray}, S.~S., {Jones}, C., {Forman}, W.,

\bibitem[{{Williamson} {et~al.}(2011){Williamson}, {Benson}, {High},
  {Vanderlinde}, {Ade}, {Aird}, {Andersson}, {Armstrong}, {Ashby}, {Bautz},
  {Bazin}, {Bertin}, {Bleem}, {Bonamente}, {Brodwin}, {Carlstrom}, {Chang},
  {Clocchiatti}, {Crawford}, {Crites}, {de Haan}, {Desai}, {Dobbs}, {Dudley},
  {Fazio}, {Foley}, {Forman}, {Garmire}, {George}, {Gladders}, {Gonzalez},
  {Halverson}, {Holder}, {Holzapfel}, {Hoover}, {Hrubes}, {Jones}, {Joy},
  {Keisler}, {Knox}, {Lee}, {Leitch}, {Lueker}, {Luong-Van}, {Marrone},
  {McMahon}, {Mehl}, {Meyer}, {Mohr}, {Montroy}, {Murray}, {Padin}, {Plagge},
  {Pryke}, {Reichardt}, {Rest}, {Ruel}, {Ruhl}, {Saliwanchik}, {Saro},
  {Schaffer}, {Shaw}, {Shirokoff}, {Song}, {Spieler}, {Stalder}, {Stanford},
  {Staniszewski}, {Stark}, {Story}, {Stubbs}, {Vieira}, {Vikhlinin}, \&
  {Zenteno}}]{williamson2011}
{Williamson}, R., et~al.  2011, arXiv:1101.1290

\bibitem[{{Yan} {et~al.}(1998){Yan}, {Sadeghpour}, \& {Dalgarno}}]{yan1998}
{Yan}, M., {Sadeghpour}, H.~R., \& {Dalgarno}, A. 1998, \apj, 496, 1044

\bibitem[{{Zeldovich} \& {Sunyaev}(1969)}]{zeldovich1969}
{Zeldovich}, Y.~B. \& {Sunyaev}, R.~A. 1969, \apss, 4, 301

\bibitem[{{Zhang} {et~al.}(2008){Zhang}, {Finoguenov}, {B{\"o}hringer},
  {Kneib}, {Smith}, {Kneissl}, {Okabe}, \& {Dahle}}]{zhang2008}
{Zhang}, Y.-Y., {Finoguenov}, A., {B{\"o}hringer}, H., {Kneib}, J.-P., {Smith},
  G.~P., {Kneissl}, R., {Okabe}, N., \& {Dahle}, H. 2008, \aap, 482, 451

\end{thebibliography}
\bibliographystyle{apj}
}

%%%%%%%%%%%%%%%%%%%%%%%%%%%%%%%%%%%%%%%%%%%%%%%%%%%%%%%%%%%%%%%%%%%%
%  Figures
%%%%%%%%%%%%%%%%%%%%%%%%%%%%%%%%%%%%%%%%%%%%%%%%%%%%%%%%%%%%%%%%%%%%
%\clearpage
%\input{fig1}
%\input{fig2}
%\input{fig3}
%\input{fig4}
%\input{fig5}

%%%%%%%%%%%%%%%%%%%%%%%%%%%%%%%%%%%%%%%%%%%%%%%%%%%%%%%%%%%%%%%%%%%%
%  Tables
%%%%%%%%%%%%%%%%%%%%%%%%%%%%%%%%%%%%%%%%%%%%%%%%%%%%%%%%%%%%%%%%%%%%
%\clearpage
%\input{tab1}
%\input{tab2}
%\input{tab3}
%\input{tab4}
%\input{tab5}
%\input{tab6}
%\input{tab7}

\end{document}